%% file: main.tex
\documentclass[format=sigconf, nonacm=true, screen=true]{acmart}
\usepackage{styles/preamble-package}

\begin{document}
\fancyhead{}

\title{Reverse Engineering Structure and Semantics of Input of a Binary Executable}

\author{Seshagiri Prabhu Narasimha and Arun Lakhotia}
\affiliation{%
    \institution{University of Louisiana at Lafayette}
    \department{School for Computing \& Informatics}
    \city{Lafayette}
    \state{Louisiana}
    \postcode{70504}
    \country{USA}
}
\email{{seshagiri.narasimha1,arun.lakhotia}@louisiana.edu}
\renewcommand{\shortauthors}{Narasimha et al.}

\input{abstract}

\maketitle

\input{paper-outline}

\bibliographystyle{unsrtnat}
\balance
\bibliography{references}

\appendix
\appendixpage
\input{appendix}

\end{document}

%% file: abstract.tex
\begin{abstract}
	Knowledge of the input format of binary executables is important for finding bugs and vulnerabilities, such as generating data for fuzzing or manual reverse engineering. This paper presents an algorithm to recover the structure and semantic relations between fields of the input of binary executables using dynamic taint analysis. The algorithm improves upon prior work by not just partitioning the input into consecutive bytes representing values but also identifying syntactic components of structures, such as atomic fields of fixed and variable lengths, and different types of arrays, such as arrays of atomic fields, arrays of records, and arrays with variant records. It also infers the semantic relations between fields of a structure, such as count fields that specify the count of an array of records or offset fields that specify the start location of a variable-length field within the input data. The algorithm constructs a C/C++-like structure to represent the syntactic components and semantic relations.

	The algorithm was implemented in a prototype system named \byteritwo. The system was evaluated using a controlled experiment with synthetic subject programs and real-world programs. The subject programs were created to accept a variety of input formats that mimic syntactic components and selected semantic relations found in conventional data formats, such as PE, PNG, ZIP, and CSV. The results show that \byteriitwo correctly identifies the syntactic elements and their grammatical structure, as well as the semantic relations between the fields for both synthetic subject programs and real-world programs. The recovered structures, when used as a generator, produced valid data that was acceptable for all the synthetic subject programs and some of the real-world programs.
\end{abstract}

%% file: paper-outline.tex
\input{introduction}

\input{background}

\input{method}

\input{implementation}

\input{evaluation}

\input{related-works}

\input{limitations}

\input{impact}

\input{conclusion}

%% file: introduction.tex
\section{Introduction}
\label{sec:introduction}

What is the structure of the valid inputs of a program? Answer to this question can help address several problems in the context of reverse engineering, vulnerability detection, migrating legacy documents, and more. For instance, it may be used to generate valid and invalid inputs to increase path coverage during fuzzing~\cite{zhu_fuzzing_survey_2022}. It may be used to generate data with and without certain features to perform differential analysis to identify code that implements a feature\cite{nagra2009surreptitious}. It may be used to create a defensive shield to filter invalid input when exposing legacy programs to the internet~\cite{allison2022progress}.

While it may appear at first glance that the aforementioned question can be addressed from program documentation, that is not always so even when a program is documented and even when the input is expected to conform to published international standards. That is so because implementations do not always conform to standards. It is not uncommon for applications to extend documented standards, either by design or as a side effect of design choices. Besides, programs and documentation rarely stay in sync. And when the documentation is available and current, it may not be in machine-processable form to be used to programmatically answer application-specific questions.

This paper continues upon a series of research that aims at answering the above question by taint-tracking, that is running the subject program with one or more inputs, observing how the program consumes the input, and recovering the structure from the resulting trace~\cite{caballero_polyglot_2007,wondracek_2008,cui_tupni_2008,hoschele2017mining,gopinath_2020,narasimha2022}. Early works~\cite{caballero_polyglot_2007,wondracek_2008} recovered structures of network protocols. Since these protocol formats have binary-encoded fixed-length fields, these early works discovered a simple, fixed-length flat structure. This line of research was extended to find simple arrays of fixed-length records~\cite{cui_tupni_2008}, but under the condition that the program accessed the array elements in either ascending or descending order within a loop. Almost a decade later, the research shifted to discovering structures of ASCII-encoded data~\cite{hoschele2017mining,gopinath_2020}. They discovered structures that could be expressed using context-free grammars but only for programs with recursive descent parsers. Recent work~\cite{narasimha2022} have expanded this line of research to finding the structures of both, binary and ASCII encoded input data, without any constraint on the program structure or how the data is accessed. They represent the recovered structure as a recursive state machine (RSM).

The problem has also been investigated using fuzzing~\cite{sochor2023fuzzing,bastani2017synthesizing,bendrissou2022synthesizing,mathis2019parser,gopinath_2020}. These methods start with the subject program and a set of valid inputs where each input is implicitly a grammar of the program's input, albeit accepting just that input. toThe methods iterate over the set of valid grammars, merge and fuzz them to generate variations, filter out invalid grammars, and stop when a convergence point is reached or resources are exhausted. Though these methods vary in how each of the steps is performed, they have one thing in common. They are limited to recovering the structures of ASCII-encoded data.

To use as a parser or a generator of valid input of a program, it is not sufficient to have the syntactic structure of the input, as has been the focus of the prior research. One may also need semantic relations that must exist between the syntactic components for an input to be valid. For example, in the BMP file format, the total number of pixels is specified using the product of the values of height and width. Thus, a parser or generator of BMP must be aware of this relationship. Since prior works do not infer semantic relations, the data generated using the structures the recover may not be acceptable to the original programs.

This paper addresses the problems of (a) recovering the structure and (b) inferring the semantic constraints between fields of the input with both binary and ASCII-encoded data. It does so by building upon the work of~\cite{narasimha2022} by using their system~\byteri\ to develop~\byteritwo. The latter improves upon the former in the following ways:
\begin{itemize}
	\item It explicitly identifies very complex structures including deeply nested arrays, arrays of variable length records, etc. Such structures were not distinguishable in the monolithic graph structure of RSM created by \byteri.
	\item It infers semantic dependence between fields of the input, where the value of one field controls the values of another field. The previous work did not infer any dependencies.
	\item It mines the semantic dependences to find complex semantic relations between fields, as for instance the relation between the height, width, and array of pixels fields in a BMP file.
	\item It recovers the structure in a C/C++-like notation that is easier to comprehend and easier to use than \byteri's state machine.
\end{itemize}

To the best of our knowledge, no prior work recovers the semantic relations between fields in the input without resorting to concolic testing~\cite{slowinska2011,cha2012}. While concolic testing can find semantic relations with high accuracy, it has two significant limitations: path explosion problems and the high computational cost of solving complex path constraints.

Some fuzzing methods use exhaustive search~\cite{you_profuzzer_2019,li_steelix_2017} or AI~\cite{shi_aifore_2023} to identify certain semantic relationships, such as the relationship between a count field and an array, and the relationship between a record type field and a variant records, where the record type field specifies the type of record.

There have also been works~\cite{lin_autoformat_2010,lin2010,jain_tiff_2018} that identify the types of values of registers and memory locations inside a program. For, instance they can find types such as \texttt{int}, \texttt{char *}, and \texttt{struct} that are passed as parameters to known library functions, such as \texttt{strcpy}, \texttt{strlen}, and \texttt{malloc}. These methods do not find semantic constraints between multiple variables.

This paper makes the following innovations:
\begin{itemize}
	\item The construction of a C/C++-like structure, instead of the RSM generated by \byteri, is a significant innovation built upon the following two important insights.

	      \begin{itemize}
		      \item The language used in C/C++ to describe structures is not equivalent to regular expressions because there is no syntactic unit that encapsulates an array. The paper introduces a language for structures that is equivalent to regular expressions.

		      \item In prior work \byteri\ abstracted the input into a sequence of nodes, with each node representing a field of the input. It treated each field as a state of a recursive state machine and each pair of successive fields in the sequence as a transition. This led to overgeneralization by creating spurious transitions between and across nested arrays. In contrast, the present work treats the set of all fields as an alphabet and constructs a regular expression from the string of characters. This may sometimes spill the last element of an array outside, but it preserves the structures of nested arrays.
	      \end{itemize}

	\item Our use of control dependences to discover semantic constraints between data items is a second significant innovation.  Control dependences are statically computed properties between program statements~\cite{podgurski1990}. Hitherto they have been used primarily for solving problems in compilers and static analysis tools, such as program slicing~\cite{harman2001overview}. On the other hand, semantic constraints between data have so far been inferred by constructing path expressions and solving those expressions, such as in concolic analysis~\cite{slowinska2011,cha2012}. Our projection of control dependences from a relation between program statements to a relation between components of data is a significant and non-trivial insight that significantly constrains the search space to mine for semantic relationships. It has enabled the discovery of rich and complex semantic relations at a fraction of the cost than that taken by creating and solving path expressions.
\end{itemize}

To gauge the accuracy of~\byteritwo, we have evaluated it using a controlled experiment with synthetic subject programs whose inputs mimic specific types of structures found in conventional data formats, such as \texttt{CSV} and \texttt{PE}, and a set of real-world applications, such as  \texttt{unzip}, \texttt{bmp\_parser}, and \texttt{pngcheck}. The recovered structures were evaluated using two methods a) manual comparison with ground truth and b) validating whether the data generated using the recovered structure is acceptable by the original subject program. Per manual comparison, the results show that our system correctly recovers the structure and semantic relations between fields of the various data formats. The data generated using the structures recovered for all the synthetic programs were valid. For real-world programs, the data generated for \texttt{unzip} and \texttt{bmp\_parser} was valid. However, even though the structure and semantic relations recovered for \texttt{pngcheck} were correct, due to over-generalization of the type of data in certain control fields, the generated data was not valid.

In all our evaluations we have used valid input data. Let that not be seen to imply that the proposed method requires only valid inputs for the program being analyzed. \byteritwo\ has no prior knowledge of the correct behavior of a program, hence it is agnostic of whether an input is valid or invalid. All that \byteritwo\ does is track the input bytes as they move through the program and abstract that into a structure. Thus, should a program determine an input to be invalid and exit, \byteritwo\ will discover the part of the input that was processed prior to determining validity and return its structure.

As stated earlier, while the problem ``What is the structure of a program input?'' has many applications, such as fuzzing, the problem is an important research problem in itself, as is evident from the series of prior research. We consider the demonstration of the method presented in aiding such other applications as a separate issue and outside the scope of assessment needed to evaluate this research.

The rest of the paper is organized as follows: Section~\ref{sec:background} introduces the terminologies and primitives used in the algorithm. Section~\ref{sec:algorithm} presents the algorithm itself, while Section~\ref{sec:implementation} provides details about its implementation. Section~\ref{sec:evaluation} describes the evaluation methods and presents the experimental results, Section~\ref{sec:related-works} discusses relevant existing research efforts, Section~\ref{sec:limitations} discusses the limitations of the presented approach, and Section~\ref{sec:Impact} describes the anticipated impact of this research. Finally, Section~\ref{sec:conclusion} concludes the paper and describes potential future research directions.

%% file: background.tex
\section {Grammar of Input}
\label{sec:background}

\input{background/introduction}

\input{background/fields-values}

\input{background/byteri-struct}

\input{background/semantic-relations}

%% file: background/introduction.tex
This section introduces a grammar to define a program's input using the language notations classically used to define internal data structures in programming languages like C/C++, as is done by reverse engineering tools such as 010~Editor~\cite{010editor}. The language notation is augmented to describe semantic relationships between data structure components thereby increasing its expressive power to create parsers and generators for complex data formats such as BMP, TIFF, ZIP, and more.

%% file: background/fields-values.tex
\subsection{Serialized Data}
\label{prelim:fields-values}

\input{tables/bmp-format}
A program takes an input--a sequence of bytes--from an external source, parses it to extract the information encoded in the bytes, and represents that information in an internal data structure. Flipping the relationship, inputs may equivalently be viewed as serialized representations of the internal data structure representation of the information extracted by a program. 

A data structure may be serialized in many ways. The choice of serialization representation depends on the purpose. When the data is to be exchanged between programs written by different entities, the serialization follows a mutually agreed standard, such as BMP, ZIP, TIFF, PDF, EXE, CSV, JSON, XML, YAML, INI, etc.

Broadly speaking, serialization formats may be classified as either binary-coded or ASCII-coded. The \emph{binary-coded data} (BCD) formats are compact and fast to deserialize. In these formats, the atomic units, such as those encoding numeric values, are of fixed size and can be ingested by a program without parsing. Thus, composite structures representing records of atomic units are also of fixed sizes. When encoding variable length data, BCD formats may encode the dimensions of the data in other atomic units. 

In contrast, \emph{ASCII-Coded Data} (ACD) formats encode the data as ASCII text. Common examples of ACD formats are CSV, JSON, XML, YAML, INI, and more. In these notations atomic fields are not commonly of fixed lengths. Thus, these notations use lexical notations to demarcate field boundaries, as also the grouping of fields into records and arrays.

Unifying the two encoding formats, this paper considers lexical notations used in ACDs as values of fields as well. Thus, commas (`,') and newlines ('$\backslash$n') in CSV format are considered values of atomic fields and serialized data, whether BCD and ACD, is at the first abstraction simply sequences of values of fields. Thus, the recovered structures for both BCD and ACD can be represented using C/C++ like structures.

%% file: tables/bmp-format.tex
\begin{figure}
	\begin{mdframed}
		\ttfamily
		\begin{adjustbox}{width=1.0\columnwidth, center}
			\begin{tabular}{@{}*{17}c@{}}
				                                              & \multicolumn{2}{c}{\scriptsize Signature} & \multicolumn{4}{c}{\scriptsize {File Size}}           & \multicolumn{4}{c}{\scriptsize {Reserved}}              & \multicolumn{4}{c}{\scriptsize {Offset of Image Data}} & \multicolumn{2}{c}{\scriptsize {Header Size}}
				\\
				000h:                                         & \cellcolor{ACMRed}{42}                    & \cellcolor{ACMRed}{4D}                                & \cellcolor{ACMBlue}{\textcolor{white}{54}}              & \cellcolor{ACMBlue}{\textcolor{white}{00}}             & \cellcolor{ACMBlue}{\textcolor{white}{00}}    & \cellcolor{ACMBlue}{\textcolor{white}{00}} & \cellcolor{gray!40!white}{00} & \cellcolor{gray!40!white}{00} & \cellcolor{gray!40!white}{00} & \cellcolor{gray!40!white}{00} & \cellcolor{ACMOrange}{36}                    & \cellcolor{ACMOrange}{00}                    & \cellcolor{ACMOrange}{00}                      & \cellcolor{ACMOrange}{00}                      & \cellcolor{gray!40!white}{28} & \cellcolor{gray!40!white}{00}
				\\
				                                              & \multicolumn{2}{c}{}                      & \multicolumn{4}{c}{\scriptsize Width}                 & \multicolumn{4}{c}{\scriptsize {Height}}                & \multicolumn{2}{c}{\scriptsize {\# Planes}}            &
				\multicolumn{2}{c}{\scriptsize {Color Depth}} &
				\multicolumn{2}{c}{\scriptsize {Compression}}
				\\
				010h:                                         & \cellcolor{gray!40!white}{00}             & \cellcolor{gray!40!white}{00}                         & \cellcolor{ACMYellow}{05}                               & \cellcolor{ACMYellow}{00}                              & \cellcolor{ACMYellow}{00}                     & \cellcolor{ACMYellow}{00}                  & \cellcolor{ACMLightBlue}{02}  & \cellcolor{ACMLightBlue}{00}  & \cellcolor{ACMLightBlue}{00}  & \cellcolor{ACMLightBlue}{00}  & \cellcolor{ACMPurple}{\textcolor{white}{01}} & \cellcolor{ACMPurple}{\textcolor{white}{00}} & \cellcolor{ACMDarkBlue}{\textcolor{white}{18}} & \cellcolor{ACMDarkBlue}{\textcolor{white}{00}} & \cellcolor{ACMGreen}{00}      & \cellcolor{ACMGreen}{00}
				\\
				                                              & \multicolumn{2}{c}{}                      & \multicolumn{4}{c}{\scriptsize {Bytes of Image}}      & \multicolumn{4}{c}{\scriptsize {Horizontal Resolution}} & \multicolumn{4}{c}{\scriptsize {Vertical Resolution}}  &
				\multicolumn{2}{c}{\scriptsize {\# Colors}}
				\\
				020h:                                         & \cellcolor{ACMGreen}{00}                  & \cellcolor{ACMGreen}{00}                              & \cellcolor{gray!40!white}{0B}                           & \cellcolor{gray!40!white}{00}                          & \cellcolor{gray!40!white}{00}                 & \cellcolor{gray!40!white}{00}              & \cellcolor{gray!50!white}{13} & \cellcolor{gray!50!white}{0B} & \cellcolor{gray!50!white}{00} & \cellcolor{gray!50!white}{00} & \cellcolor{gray!40!white}{13}                & \cellcolor{gray!40!white}{0B}                & \cellcolor{gray!40!white}{00}                  & \cellcolor{gray!40!white}{00}                  & \cellcolor{olive}{00}         & \cellcolor{olive}{00}
				\\
				                                              & \multicolumn{2}{c}{}                      & \multicolumn{4}{c}{\scriptsize {\# Important colors}} & \multicolumn{3}{c}{\scriptsize {Pixel \#1}}             & \multicolumn{3}{c}{\scriptsize {Pixel \#2}}            & \multicolumn{3}{c}{\scriptsize {Pixel \#3}}
				\\
				030h:                                         & \cellcolor{olive}{00}                     & \cellcolor{olive}{00}                                 & \cellcolor{pink}{00}                                    & \cellcolor{pink}{00}                                   & \cellcolor{pink}{00}                          & \cellcolor{pink}{00}                       & \cellcolor{gray!20!white}{00} & \cellcolor{gray!20!white}{00} & \cellcolor{gray!20!white}{FF} & \cellcolor{gray!5!white}{FF}  & \cellcolor{gray!5!white}{FF}                 & \cellcolor{gray!5!white}{FF}                 & \cellcolor{gray!20!white}{FF}                  & \cellcolor{gray!20!white}{00}                  & \cellcolor{gray!20!white}{00} & \cellcolor{gray!5!white}{00}
				\\ \multicolumn{1}{c}{} & \multicolumn{2}{c}{\scriptsize {Pixel \#4}} & \multicolumn{3}{c}{\scriptsize {Pixel \#5}} & \multicolumn{3}{c}{\scriptsize {Pixel \#6}} & \multicolumn{3}{c}{\scriptsize {Pixel \#7}}  & \multicolumn{3}{c}{\scriptsize {Pixel \#8}} & \multicolumn{2}{c}{\scriptsize {Pixel \#9}}
				\\
				040h:                                         & \cellcolor{gray!5!white}{FF}              & \cellcolor{gray!5!white}{00}                          & \cellcolor{gray!20!white}{FF}                           & \cellcolor{gray!20!white}{00}                          & \cellcolor{gray!20!white}{00}                 & \cellcolor{gray!5!white}{FF}               & \cellcolor{gray!5!white}{FF}  & \cellcolor{gray!5!white}{FF}  & \cellcolor{gray!20!white}{FF} & \cellcolor{gray!20!white}{00} & \cellcolor{gray!20!white}{00}                & \cellcolor{gray!5!white}{FF}                 & \cellcolor{gray!5!white}{FF}                   & \cellcolor{gray!5!white}{FF}                   & \cellcolor{gray!20!white}{00} & \cellcolor{gray!20!white}{00}
				\\ \multicolumn{2}{c}{} & \multicolumn{3}{c}{\scriptsize {Pixel \#10}}
				\\
				050h:                                         & \cellcolor{gray!20!white}{00}             & \cellcolor{gray!5!white}{FF}                          & \cellcolor{gray!5!white}{FF}                            & \cellcolor{gray!5!white}{00}
			\end{tabular}
		\end{adjustbox}
	\end{mdframed}
	\captionsetup{justification=centering, font=small}
	\captionof{figure}[bmp_bcd]{Binary-coded 5x2 pixel BMP image.}
	\label{fig:bmp-bcsd}
	\Description{}{}
\end{figure}

%% file: background/byteri-struct.tex
\subsection{Language of Structure}
\label{prelim:struct}

\input{diagrams/token}
\input{diagrams/structure}

The language used by~\byteriitwo to represent the recovered grammar are presented in Figures~\ref{fig:byteri-token},~\ref{fig:byteri-structure}, and~\ref{fig:byteri-semantics}. The language provides for describing the byte pattern of data in atomic fields (Figure~\ref{fig:byteri-token}), for describing the composite structure of the encoded data structure (Figure~\ref{fig:byteri-structure}), and for describing the semantic relationships between the components (Figure~\ref{fig:byteri-semantics}).  Though this notation demonstrates the ability to infer these relationships by methods used by \byteritwo, there is no claim that the notation is sufficient to express the entirety of relations in some complex data structures such as that may be described using the Template language of 010~Editor~\cite{010editor}.

\paragraph{Token.}
The language described by \texttt{Token}, Figure~\ref{fig:byteri-token}~\cite{narasimha2022}, consists of regular expressions over the alphabet \texttt{Unit}, where each symbol of \texttt{Unit} is a regular expression that matches specific byte values. The symbols 0x00 to 0xFF match the corresponding 8-bit binary values (256 values). The other symbols match one of a subset of binary values. For example, the symbol DIGIT matches any byte value in the range 0x30 to 0x39, the ASCII range of `0' to `9'. A regular expression that does not terminate with a `$+$' matches byte sequences of the same length as the regular expression and whose $i^{th}$ byte belongs to the set of bytes represented by the regular expression's $i^{th}$ symbol. A regular expression terminating with a `$+$' matches any byte sequence that matches a regular expression in which the symbol preceding `$+$' is repeated one or more times. Thus, the regular expression \texttt{[DIGIT DIGIT]} matches a two-length sequence consisting of byte values in the range 0x30 and 0x39, and the regular expression \texttt{[DIGIT +]} matches any finite length sequence of one or more bytes from the range 0x30 to 0x39.

\paragraph{Structure.}
The language described by \emph{Structure}, Figure~\ref{fig:byteri-structure}, consists of a grammar that describes the structure of inputs in a C/C++-like notation. An element of \texttt{Structure} is either an \texttt{atomic} `field' or a composite `field'. A composite field---a composition of atomic and composite fields---may be either an \texttt{array}, a \texttt{record}, or an \texttt{option} field.

Each field describes a grammar that accepts a certain sequence of bytes as follows:

\justify{}
\begin{itemize}
    \item \texttt{\textbf{atomic} id = t}, where \texttt{id} $\in$ \texttt{Id} and \texttt{t} $\in$ \texttt{Token}: Accept a byte sequence accepted by \texttt{t}.

    \item \texttt{\textbf{array} id f}, where \texttt{id} $\in$ \texttt{Id} and \texttt{f} $\in$ \texttt{Structure}: Accepts a sequence of one or more byte sequences accepted by \texttt{f}.

    \item \texttt{\textbf{record} id [f$_1$,$\ldots$,f$_n$]}, where \texttt{id} $\in$ \texttt{Id} and $\forall_{i\in\{1,\ldots,n\}}$ \texttt{f$_i$} $\in$ \texttt{Structure}: Accept a byte sequence \texttt{[b$_1$,$\ldots$,b$_n$]}, where \texttt{b$_i$} is a byte sequence accepted by \texttt{f$_i$}.

    \item \texttt{\textbf{option} id [f$_1$,$\ldots$,f$_n$]}, where \texttt{id} $\in$ \texttt{Id} and $\forall_{i\in\{1,\ldots,n\}}$ \texttt{f$_i$} $\in$ \texttt{Structure}: Accepts a byte sequence \texttt{b$_i$}, where \texttt{b$_i$} is a byte sequence accepted by \texttt{f$_i$}.
\end{itemize}

The language notation, though, varies slightly from conventional C/C++ notations in that the abstract syntax tree of data structures it generates is equivalent in shape to regular expressions, which forms the basis of \byteritwo's capability to represent the recovered data structure as a grammar, as opposed to \byteri's recursive state machine. The correspondence between the abstract syntax tree of \texttt{Structure} and regular expressions is apparent and left to the reader.

Although the grammars of the several formats, such as JSON and XML, are context-free, it is not necessary that the data encoded using such formats are also context free. For instance, if the data encoded in XML format doesn't contain nesting of structures of the same type, as in binary tries, it can be represented as regular expression with the XML keywords and delimeters encoded as fields and can be represented using the language notation introduced here in.

%% file: diagrams/token.tex
\begin{figure}
\begin{mdframed}
\begingroup
\ttfamily\scriptsize
\begin{bnf*}
    \bnfprod{Token}{\bnfts{List}\bnfpn{Unit}\bnfts{[+]}}\\
    \bnfprod{Unit}{\bnfts{0x0} \bnfor \bnfsk \bnfor \bnfts{0xFF}}\\
    \bnfmore{\bnfor \bnfts{LOWER\_HEX} \bnfor \bnfts{UPPER\_HEX} \bnfor
        \bnfts{LOWER} \bnfor \bnfts{UPPER}}\\
    \bnfmore{\bnfor \bnfts{DIGIT} \bnfor \bnfts{XDIGIT}
        \bnfor \bnfts{ALPHA} \bnfor  \bnfts{ALNUM}}\\
    \bnfmore{\bnfor \bnfts{WHITESPACE} \bnfor \bnfts{PUNCTUATION} \bnfor \bnfts{CONTROL}}\\
    \bnfmore{}{\bnfor \bnfts{PRINTABLE} \bnfor \bnfts{ALL}}\\
\end{bnf*}
\endgroup
\end{mdframed}
\captionsetup{justification=centering, font=small}
\caption{Meta-grammar \texttt{Token}~\cite{narasimha2022}}
\label{fig:byteri-token}
\Description{}{}
\end{figure}

%% file: diagrams/structure.tex
\begin{figure}
\begin{mdframed}
\begingroup
\ttfamily\scriptsize
\begin{bnf*}
	\bnfprod{Structure}{\bnfts{\textbf{atomic} } \bnfpn{Id} \bnfts{ = } \bnfpn{Token} \bnfts{ } \bnfpn{Constraints}}\\
	\bnfmore{\bnfor \bnfts{\textbf{array} } \bnfpn{Id} \bnfts{ } \bnfts{\{} \bnfpn{Structure} \bnfts{\}} \bnfts{ } \bnfpn{Constraints}}\\
	\bnfmore{\bnfor \bnfts{\textbf{record} } \bnfpn{Id} \bnfts{ } \bnfts{\{} \bnfts{List}\bnfpn{Structure} \bnfts{\}} \bnfts{ } \bnfpn{Constraints}}\\
	\bnfmore{\bnfor \bnfts{\textbf{option} } \bnfpn{Id} \bnfts{ } \bnfts{\{} \bnfts{List} \bnfpn{Structure} \bnfts{\}} \bnfts{ } \bnfpn{Constraints}}\\
\end{bnf*}
\endgroup
\end{mdframed}
\captionsetup{justification=centering, font=small}
\caption{Meta-grammar \texttt{Structure}}
\label{fig:byteri-structure}
\Description{}{}
\end{figure}

%% file: background/semantic-relations.tex
\subsection{Semantic Relations}
\label{prelim:semantic}
\input{diagrams/semantic_rules}

As described earlier, more than just a grammar describing the structure of byte sequences is required to use as a parser or a generator. The grammar must also be augmented with semantic constraints that bind different fields. For example, in the BMP format, Figure~\ref{fig:bmp-bcsd}, the number of \texttt{Pixels} fields is equal to the product of the integer values in \texttt{Width} and \texttt{Height} fields.

Figure~\ref{fig:byteri-semantics} presents a language notation to attach semantic relations to fields generated by \texttt{Structure}, Figure~\ref{fig:byteri-structure}. The language uses C/C++ notation to reference a specific field, such as \texttt{f1}, \texttt{f1.f2}, \texttt{f1[i]}, or a composition of them. The relations are described in terms of the following properties:
\begin{itemize}
	\item \texttt{f.\textbf{bytes}}: The byte sequence of field \texttt{f}. When \texttt{f} is a composite field, \texttt{f.\textbf{bytes}} represents the byte sequence for the entire composite structure.

	\item \texttt{f.\textbf{offset}}: The \emph{start offset} of the first byte of the field \texttt{f} relative to the entire data starting from the root id of \texttt{f}.

	\item \texttt{f.\textbf{size}}: The \emph{number of bytes} in \texttt{f.\textbf{bytes}}.

	\item \texttt{f.\textbf{terminator}}: Relevant when \texttt{f} identifies a variable length \texttt{\textbf{atomic}} field. It represents the sequence of bytes that follow and mark the termination of \texttt{f.\textbf{bytes}}.

	\item \texttt{f.\textbf{count}}: Relevant when \texttt{f} references an \texttt{\textbf{array}}; represents the \emph{number of values} in \texttt{f}.

	\item \texttt{f.\textbf{record\_type}}: Relevant when the root of \texttt{f} is an \texttt{\textbf{option}} structure, say \texttt{o}, and the value of field \texttt{f} serves to distinguish between the different structures that are accepted by \texttt{o}.
\end{itemize}

The size of an atomic field is determined by the sizes of the byte sequences accepted by the associated token. A field is fixed-length if all the accepted byte sequences are the same length. Otherwise, the field is variable length with the minimum length equal to the shortest byte sequence accepted.

Additionally, the function \texttt{\textbf{int()}} represents the integer value (\texttt{f.\textbf{bytes}}) when \texttt{f} is an atomic field.

With the described properties, one can define relationships between fields. The language facilitates describing more complex constraints required for the grammar of ZIP, TIFF, PE, and other complex data formats. An example of representing the structure of BMP image is available in Appendix 
\ref{appendix:use-case-bmp}.

%% file: diagrams/semantic_rules.tex
\begin{figure}
\begin{mdframed}
\begingroup
\ttfamily\scriptsize
\begin{bnf*}
    \bnfprod{Constraints}{\bnfts{WHERE } \bnfpn{Relation} \bnfts{ (} \bnfts{OR } \bnfpn{Relation} \bnfts{)* [}\bnfpn{Constraints} \bnfts{]}}\\
    \bnfprod{Relation}{\bnfpn{Acc} \bnfts{.} \bnfpn{IntProperty} \bnfts{ = } \bnfpn{IntValue}}\\
    \bnfmore{\bnfor \bnfpn{Acc} \bnfts{.} \bnfpn{BytesProperty} \bnfts{ = } \bnfpn{Acc} \bnfts{\textbf{.bytes}}}\\
    \bnfprod{IntProperty}{\bnfts{\textbf{count}} \bnfor \bnfts{\textbf{size}} \bnfor \bnfts{\textbf{offset}}}\\
    \bnfprod{BytesProperty}{\bnfts{\textbf{record\_type}} \bnfor \bnfts{\textbf{terminator}}}\\
    \bnfprod{IntValue}{\bnfts{\textbf{int(}} \bnfpn{Acc} \bnfts{\textbf{.bytes}}\bnfts{\textbf{)} [} \bnfts{(- } \bnfor \bnfts{ +)}\bnfts{ 1} \bnfts{]}}\\
    \bnfprod{Acc}{\bnfpn{Id} \bnfor \bnfpn{Acc}[\bnfpn{Var}] \bnfor \bnfpn{Acc}.\bnfpn{Id}}\\
\end{bnf*}
\endgroup
\end{mdframed}
\captionsetup{justification=centering, font=small}
\caption{Semantic rules of \texttt{Structure}}
\label{fig:byteri-semantics}
\Description{}{}
\end{figure}

%% file: byteri-semantics-paper/Method.tex
\section{Algorithm}
\label{sec:algorithm}

\input{method/problem}

\input{method/overview}

\input{method/trace-to-field-trace}

\input{method/construct-structure}

\input{method/semantic-dependent-fields}

\input{method/infer-semantics}

%% file: method/problem.tex
Any solution to the problem under investigation would involve solving the following:
\begin{enumerate}[label=(\roman*)]
    \item \textbf{Partition input into fields.} Identify the values and fields in the input. This requires solving the following:
          \begin{enumerate}
              \item \textbf{Identify values.}
                    Split the input, which is a sequence of bytes, into a smaller sequence representing values. This converts the input into a sequence of values~\cite{narasimha2022}.
              \item \textbf{Identify fields.}
                    Group values belonging to the same field, such as a field in an array of records. This converts the sequence of values into a sequence of fields~\cite{narasimha2022}.
          \end{enumerate}
    \item \textbf{Construct structure.}
          Translate the sequence of fields to an element of \texttt{Structure}.
    \item \textbf{Identify semantic dependence between fields.}
          This requires solving the following:
          \begin{enumerate}
              \item \textbf{Identify control-dependent statements.}
                    Identify\\control-dependent statements of a program where the behavior of a statement affects the execution behavior of another statement, such as a branch statement that determines the number of times the body of the loop that it controls is executed.
              \item \textbf{Identify control-dependent fields.}
                    Identify fields used by control-dependent statements, such as a counter field used at a branch statement of a loop that processes an array.
          \end{enumerate}
    \item \textbf{Infer semantic relations.}
          Identify semantic constraints between fields, such as a count field that specifies the count of an array of records.
\end{enumerate}

%% file: method/overview.tex
\input{diagrams/algorithm-steps}

The proposed algorithm takes four steps to infer the semantic relations between fields in the input, as shown in Figure~\ref{fig:algorithm-steps}.

Given a well-formed input and an arbitrary program that processes the input data, the algorithm first converts the sequence of bytes in the input to a sequence of fields. Subsequently, it converts the sequence of fields to an element of \texttt{Structure}. Next, it computes the control dependencies between the statements of the program. Following that, it identifies fields with semantic dependencies. Finally, the algorithm infers the semantic relationships between the components of the structure, generating a grammar that encapsulates the structure and semantic connections between its components. For a running example of the algorithm, see the Appendix~\ref{appendix:running-example}.

The following subsections include the details of each step of the algorithm.

%% file: diagrams/algorithm-steps.tex
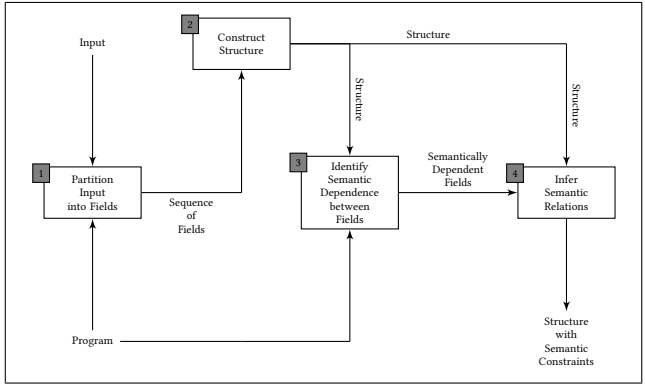
\begin{figure}
	\centering
	\begin{mdframed}
		\begin{tikzpicture}[
				scale=0.72, transform shape
			]
			\tikzset{
				->,
				>=stealth',
				node distance=2.75cm,
				auto,
				file/.style={
						rectangle,
						draw=black,
						text width=5em,
						align=center,
						minimum height=3em,
						label={
								[draw, rectangle, fill=gray, xshift=-0.95cm, yshift=-0.30cm]
								#1
							}
					},
				textbox/.style={
						text width=5em,
						align=center,
					},
				textbox_small/.style={
						text width=2.5em,
						align=center,
					},
				input/.style  = {coordinate},
				font=\scriptsize,
			}
			\node [file=1] (byteri) {Partition\\Input\\into Fields};
			\node [textbox, above of=byteri] (input) {Input};
			\node [textbox_small, below of=byteri] (program) {Program};

			\node [input, right of=byteri] (byteri_point) {};
			\node [file=2, above of=byteri_point] (structure) {Construct\\Structure};
			\node [input, below of=byteri_point] (control_dependence) {};
			\node [file=3, right of=byteri_point, node distance=2cm] (semantic_dependence) {Identify\\Semantic\\Dependence\\between\\Fields};
			\node [file=4, right of=semantic_dependence, node distance=4cm] (semantic_constraints) {Infer\\Semantic\\Relations};
			\node [textbox, below of=semantic_constraints] (structure_contraints) {Structure\\with\\Semantic\\Constraints};

			\node [input, right of=structure, node distance=2cm] (structure_point) {};
			\node [input, above of=semantic_constraints] (semantic_point) {};
			\node [input, right of=control_dependence, node distance=2cm] (control_dependence_point) {};
			\node [input, right of=semantic_dependence] (semantic_dependence_point) {};

			\path[line] (input) -- node[] {} (byteri);
			\path[line] (program) -- node[] {} (byteri);
			\path[line, -] (program) -- node[] {} (control_dependence);
			\path[line, -] (byteri) -- node[below, align=center] {Sequence\\of\\Fields} (byteri_point);
			\path[line] (byteri_point) -- node[left] {} (structure);

			\path[line, -] (structure) -- node [above]{} (structure_point);
			\path[line, -] (structure) -- node [above]{Structure} (semantic_point);
			\path[line, -] (control_dependence) -- node [above, align=center]{} (control_dependence_point);

			\path[line] (structure_point) -- node[above, rotate=-90] {Structure} (semantic_dependence);
			\path[line] (semantic_point) -- node[above, rotate=-90] {Structure} (semantic_constraints);
			\path[line] (control_dependence_point) -- node[right, align=center] {} (semantic_dependence);
			\path[line] (semantic_dependence) -- node[above, align=center] {Semantically\\Dependent\\Fields} (semantic_constraints);
			\path[line] (semantic_constraints) -- node[] {} (structure_contraints);
		\end{tikzpicture}
	\end{mdframed}
	\captionsetup{justification=centering, font=small}
	\caption{Steps to infer semantic constraints between fields}
	\label{fig:algorithm-steps}
	\Description{}{}
\end{figure}

%% file: method/trace-to-field-trace.tex
\subsection{Partition Input into Fields}
\label{sec:partition-input-fields}

The algorithm relies on a dynamic taint tracking tool. We use the dynamic taint tracker of \byteri~\cite{narasimha2022}, and improved the tracker by adding these two features:
\begin{itemize}
	\item It groups contiguous tainted bytes that reach an instruction and propagates them as a single unit.
	\item It tracks an abstracted calling context in which two active (recursive) calls from the same call site are considered to have the same context. Otherwise, each unique sequence of active calls will have a unique context.
\end{itemize}

The abstracted calling context distinguishes between two executions of the same instruction in the absence of recursion. However, if the same instruction is executed multiple times in a recursive loop, it is considered to be executed in the same calling context.

In a recursive call, the call site of the first call is different from the call sites of subsequent calls. For example, the call site $g.sx$ calls function $f$, which then calls $f$ recursively at call site $f.sy$. In this case, the call site of the first call at $g.sx$ is different from the call site of the subsequent calls at $f.sy$. The context is truncated at the first call site in order to have recursive calls to $f$ have the same calling context. Otherwise, the calling context at the first call would be different from the calling contexts of the subsequent calls to $f$.

The taint tracker produces a \textit{taint trace}, which is a sequence of tuples of the form $(<I, C, T>)$. Each tuple has an instruction address $I$, a truncated calling context $C$, and a set of taints $T$. The taints $T$ are the set of values or taint intervals used by an instruction during an invocation of that instruction.

Using the taint trace and the two heuristics defined in~\cite{narasimha2022}, we identify:
\begin{itemize}
	\item \textbf{Values}. A \textit{value} of a taint interval is a contiguous range of bytes in the input data that is tainted and is used by at least one instruction~\cite{narasimha2022}.
	\item \textbf{Fields}. A \textit{field} is a set of values used by exactly the same set of instructions, albeit during different invocations~\cite{narasimha2022}.
\end{itemize}

A field is a conceptual entity in a program that is not directly represented in the input data. A field is a set of instruction addresses in the program, along with the calling context where its values are used during execution. Each field is assigned an identifier called the source index (SI), the set of all the instruction addresses, paired with the calling contexts ($(<I, C>)$), at which the values of a field are used~\cite{narasimha2022}. Some fields in the input may only have one value, while others may have two or more values. Fields with multiple values represent repeating structures in the input, such as arrays. In that case, all the values of a field in the array will have the same SI.

Since a taint interval forms a value and one or more values form a field, a taint interval could be mapped to a field. After identifying the values and fields from the taint trace, replace references to taint intervals in the taint trace $(<I, C, T>)$ with their corresponding fields $F$. This produces a sequence of tuples of the form $(<I, C, F>)$.

The \emph{Taint Interval Graph} (TIG) is constructed by starting with the set of taint intervals from the taint trace and adding an interval that spans the entire input. The transitive reduction of the subset relation of the taint intervals is then taken to construct the TIG. To convert a TIG to a TIG tree, interval gaps are filled and overlapping intervals are split.

The next step involves identifying co-occurring fields. Two fields are considered to \emph{co-occur} if they share at least one instruction (paired with its corresponding calling context) in their SIs. In other words, if field \texttt{F$_a$} is identified using a set of instruction addresses paired with the calling contexts $(I_m, C_n)$ and \texttt{F$_b$} is identified using $(I_x, C_y)$, then $(I_m, C_n)$ $\cap$ $(I_x, C_y)$ $\neq$ $\phi$.

The \texttt{New\_SI} of fields is calculated, representing the union of the instruction addresses paired with the calling contexts of the co-occurring fields. For instance, if fields \texttt{F$_a$} and \texttt{F$_b$} are co-occurring, their \texttt{New\_SI} would be $(I_m, C_n)$ $\cup$ $(I_x, C_y)$. The \texttt{New\_SI} calculation is performed for all nodes in the TIG tree. If a TIG node is the value of a field that does not co-occur with other fields, its \texttt{New\_SI} would be the same as its SI.

A \emph{frontier} of a TIG tree is a sequence of TIG nodes whose intervals collectively cover the interval of the root node without any overlaps or gaps. A TIG tree can have multiple frontiers, each depicting a distinct level of abstraction of the input.

The most intriguing values are those that belong to an array. These correspond to the TIG nodes whose values have a repeating SI. Some of these TIG nodes with repeating SIs may have descendants with repeating SIs, suggesting the presence of a nested array. Since a TIG tree can have TIG nodes with repeating SIs at various levels, it is crucial to identify all frontiers with repeating SIs at every level of the TIG tree, starting from the top and moving downwards.

Frontiers are extracted from a TIG tree by recursively applying the \texttt{topmost\_reps} algorithm~\cite{narasimha2022} to the root node of the TIG tree. This process generates a set of frontiers with increasing levels of granularity. The outcome is a map from SI to a collection of frontiers, where SI is either that of the root node of the tree or a TIG node with repeating SI. Similarly, the {topmost\_reps} algorithm~\cite{narasimha2022} is applied recursively to the root node of the TIG tree to produce a set of frontiers, in which \texttt{New\_SI} is used instead of SI. This results in a map from \texttt{New\_SI} to a collection of frontiers.

%% file: method/construct-structure.tex
\subsection{Construct Structure}
\label{sec:construct-structure}

This section outlines a method for constructing the structure of input from a frontier. This method involves mapping the problem of transforming a frontier into a structure to the equivalent problem of transforming a string into a regular expression.

The \regex algorithm~\cite{construct_regex_2024} is capable of transforming a string into a regular expression. Since a frontier represents a sequence of values or fields, the \regex algorithm can be applied to convert a frontier into a regular expression. The resulting regular expression would be equivalent to the AST of the \texttt{Structure}. In other words, the \regex algorithm produces an element of \texttt{Structure} from a frontier. While the mapping between transforming a frontier into a structure and transforming a string into a regular expression is not immediately apparent, the \regex algorithm effectively accomplishes this task.

The program's behavior can sometimes lead to inaccurate boundaries for \texttt{records} within an \texttt{array} structure. This occurs when an instruction uses the first or last element of an array separately from the instructions that access all array elements in a loop or recursion. For instance, elements 1 through $n$ of an array are used by instructions ($I_x, C_y$), while the first element is used by an additional instruction ($i, c$). This results in a different SI set for the first element (($I_x, C_y$) $\cup$ ($i, c$)) compared to the remaining elements ($I_x, C_y$) of the \texttt{array}, although they represent the same component in the structure. Consequently, the first element has a distinct SI, leading to its potential misinterpretation as a separate component outside the \texttt{array} structure.

When the recovered structure contains an \texttt{array} or an \texttt{option} nested within an \texttt{array}, a heuristic is applied to rectify the boundaries of the contained \texttt{records}. This heuristic works on the principle that all the elements of a repeating structure are accessed by \emph{mostly} the same set of instructions. If these elements are inadvertently split due to program behavior, their SIs will exhibit a significant overlap of instruction addresses (paired with calling contexts). This overlap identifies them as co-occurring fields, leading to a shared \texttt{New\_SI}. Conversely, if the elements are not split, they will have the same SI and \texttt{New\_SI}. This characteristic of the heuristic plays a vital role in fixing the boundaries of \texttt{records} nested in an \texttt{array} or \texttt{option} that are incorrectly split due to program behavior.

If the heuristic is employed, a frontier is extracted from the map associating \texttt{New\_SI} with a collection of frontiers. Conversely, if the heuristic is not employed, a frontier is extracted from the map associating SIs with a collection of frontiers. The chosen frontier is then provided as input to the \regex algorithm, which generates a new structure utilizing \texttt{New\_SI} instead of the original SIs.

%% file: method/semantic-dependent-fields.tex
\subsection{Identify Semantic Dependence between Fields}
\label{sec:semantic-dependent-fields}

\begin{figure}
    \centering
    \captionsetup{justification=centering, font=small}
\begin{lstlisting}[
        style=pystyle,
        caption={\texttt{project\_graph} -- Graph Projection Algorithm},
        label={lst:project_graph},
        morekeywords={Input, Output}
    ]
def dfs(v, visited, icdg, N):
    visited.append(v)
    reachable = set()
    for neighbor in icdg[v]:
        if neighbor in N:
            reachable.add(neighbor)
        elif neighbor not in visited:
            r_neighbors = dfs(neighbor, visited, icdg, N)
            reachable.update(r_neighbors)
    return reachable

def project_graph(icdg, N):
    projected_graph = dict(set)
    for node in N:
        visited = list()
        reachable = dfs(node, visited, icdg, N)
        projected_graph[node].update(reachable)
    return projected_graph
\end{lstlisting}
    \Description{}{}
\end{figure}

This section proposes a method for identifying semantic dependence between fields, and it does so in two steps: (a) identifying program statements with control dependence and (b) identifying fields used at control-dependent statements.

\subsubsection{Identify program statements with control dependence}
The control dependence between the statements in the program is computed using the \texttt{ComputeInterCD} algorithm~\cite{sinha01}. This algorithm takes as input the control flow graphs (CFGs) of the procedures within a binary executable as input and outputs an interprocedural control dependencies graph (ICDG).

Nodes ($N^{ICDG}$) in the ICDG graph ($G^{ICDG}$) represent basic blocks or nodes ($N^{ICFG}$) of the Interprocedural Control-Flow Graph (ICFG)~\cite{landi_1992}, while edges ($E^{ICDG}$) denote the control dependencies between them. Since this research does not focus on recovering CFGs, it relies on a dedicated CFG extractor tool~\cite{ghidra-re} to extract the CFGs from the binary executables.

Sinha \emph{et al}~\cite{sinha01} define control-dependence as follows: node $y$ $\in$ $N^{ICFG}$ is control dependent on node $x$ $\in$ $N^{ICFG}$ when node $x$ conditionally guards the execution of node $y$ ($x$ $\delta^{c}$ $y$), where the symbol $\delta^{c}$ denotes the control dependence.

Based on this definition of control dependence, we propose a heuristic to identify the semantic dependence of fields in the input. This heuristic relies on control dependence and is defined as follows:

\begin{quote}
    If a program statement $S$ has a control dependence with another statement $S'$ ($S'$ $\delta^c$ $S$), and if field $F_a$ is used at statement $S$ and field $F_{x}$ is used at statement $S'$, then field $F_a$ has semantic dependence on field $F_{x}$ ($F_{x}$ $\delta^s$ $F_{a}$).
\end{quote}

Here, the symbol $\delta^{s}$ denotes the semantic dependence.

\subsubsection{Identify fields used at control-dependent statements}
This method identifies the basic blocks where fields of the structure of the input are used. It takes the field trace ($<I, C, F>$), the structure of the input ($S$), and ICDG ($G^{ICDG}$) of the binary executable as input.

The identification process involves finding basic blocks whose instructions share at least one instruction with the instructions representing a specific field. This is achieved by checking if the intersection of the instructions of the field ($I_f$) and the instructions of the basic block ($I_{bb}$) is non-empty. This step iterates for all fields in $S$, resulting in a set $N$ of relevant basic blocks or nodes in $G^{ICDG}$, where $N$ $\subseteq$ $N^{ICDG}$. Nodes in $G^{ICDG}$ are then annotated with the fields that share instructions with the basic blocks they represent.

The algorithm for graph projection, \texttt{project\_graph}, is presented in Listing~\ref{lst:project_graph}. For each node $n$ $\in$ $N$, it performs a depth-first search to identify the reachable nodes $K$, where $K$ $\subset$ $N$. This process tracks visited nodes to avoid cycles in $G^{ICDG}$. The result is a \emph{projected graph} containing only nodes in $N$ with edges representing direct or transitive control dependencies.

A heuristic infers \emph{semantic dependence} between fields associated with nodes connected by an edge in the projected graph. If $bb_x$ is control-dependent on the guard block $bb_y$, and $F_{bb_x}$ and $F_{bb_y}$ represent the fields used by their instructions, then $F_{bb_x}$ is semantically dependent on $F_{bb_y}$ (denoted by $F_{bb_y}$ $\delta^s$ $F_{bb_x}$).

A backward traversal of the projected graph starting at a node and following directed edges back to its predecessors reveals the nodes it is control-dependent on. This sequence of nodes forms a \emph{chain of dependence}. For example, if a backward traversal from $bb_1$ ends at  $bb_k$, the sequence \{$bb_1$, $bb_2$, $\ldots$, $bb_k$\} represents the chain of dependence for $bb_1$.

Furthermore, if $F_{bb_i}$ represents the fields used by node $bb_i$ and  \{$F_{bb_j}$, $\ldots$, $F_{bb_k}$\} represents the fields used by the nodes in its chain of dependence, then \{$F_{bb_j}$, $\ldots$, $F_{bb_k}$\} forms the \emph{chain of dependent fields} for $F_{bb_i}$.

%% file: method/infer-semantics.tex
\subsection{Infer Semantic Relations}
\label{sec:semantic-constraints}

Given a pair of fields, \texttt{f$_t$} and \texttt{f$_s$} in the \texttt{Structure} ($S$), with semantic dependence (\texttt{f$_s$} $\delta^s$ \texttt{f$_t$}), the following are the different types of semantic constraints that are tested for:

\subsubsection{Size}
When a variable-length field \texttt{f$_t$}, either \atomic\ or \record, is semantically dependent on an \atomic\ field \texttt{f$_s$} and uses \texttt{f$_s$} as its size field, their relationship follows the \emph{size relation}:
\begin{equation}\label{eq:size}
    \texttt{f}_t.\textbf{size}\\ =\ \textbf{int}(\texttt{f}_s.\textbf{bytes})\ [(-\ |\ +)\ 1]
\end{equation}
Field \texttt{f$_s$} dictates the size of field \texttt{f$_t$} and is typically used to guard a loop or recursive function that processes \texttt{f$_t$}. This creates a semantic dependence between the two fields.

\subsubsection{Terminator}
When a variable length field \texttt{f$_t$}, either \atomic\ or \record, is demarcated by an \atomic\ field \texttt{f$_s$}, their relationship follows the \emph{termination relation}:
\begin{equation}\label{eq:terminator}
    \texttt{f}_t.\textbf{terminator}\ =\ \texttt{f}_s.\textbf{bytes}
\end{equation}
While variable-length fields can be marked by size or terminator fields, or both, size fields always appear before them, while terminator fields follow them in the data. Terminator fields typically hold fixed values and are often used only within library functions. For these cases, only the terminator's value matters, as the two fields may not have a semantic dependence.

\subsubsection{Count}
When an \atomic\ field contained in an \arr\ \texttt{f$_t$} is semantically dependent on another \atomic\ field \texttt{f$_s$} and uses \texttt{f$_s$} as its count field, their relationship follows the \emph{count relation}:
\begin{equation}\label{eq:count}
    \texttt{f}_t.\textbf{count}\ =\ int(\texttt{f}_s.\textbf{bytes})\ [(-\ |\ +)\ 1]
\end{equation}
Field \texttt{f$_s$} specifies the number of elements in the \arr\ and is typically used to guard a loop or recursive function that processes the \arr. This creates a semantic dependence between the two fields.

\subsubsection{Offset}
When a variable-length field \texttt{f$_t$}, either \atomic\ or \record, uses another \atomic\ field \texttt{f$_s$} as its offset field, their relationship follows the \emph{offset relation}:.
\begin{equation}\label{eq:offset}
    \texttt{f}_t.\textbf{offset}\ =\ int(\texttt{f}_s.\textbf{bytes})\ [(-\ |\ +)\ 1]
\end{equation}
The offset field determines the location of \texttt{f$_t$} within the input data. For instance, in a file, the offset field tells functions like \texttt{fseek} or \texttt{lseek} where to move the pointer to reach \texttt{f$_t$}. In such cases, identifying the fields driving such pointer movement is crucial, as the two fields may not have a semantic dependence.

\subsubsection{Record Type}
When an \arr\ \texttt{f$_t$} contains variant records (\option) and all records share a common atomic field \texttt{f$_s$} that specifies the record type, their relationship follows the \emph{enumeration relation}:
\begin{equation}\label{eq:record-type}
    \texttt{f}_t.\textbf{record\_type}\ =\ \texttt{f}_s.\textbf{bytes}
\end{equation}
In an array with variant records, each record typically has a field indicating its type. This field has a unique value for each record type. Programs processing such input often use \texttt{if} conditions or \texttt{switch} statements to check the record type before processing its remaining fields, creating a semantic dependence between \texttt{f$_s$} and the remaining fields.

In some cases, count relations can be expressed as functions of an \atomic\ field and the number of elements of an \arr\ field. This research explored input formats where count relations are simple mathematical functions, like \emph{product} or \emph{modulus}. While relations in other formats might involve more complex functions, identifying all possibilities presents an engineering challenge left for future work. This paper focuses on inferring two simple count expressions: modulus and product.

\subsubsection{Modulus}
When an \atomic\ field contained in an \arr\ \texttt{f$_t$} is semantically dependent on another \atomic\ field \texttt{f$_s$}, and the value of field \texttt{f$_s$} is a divisor of the count of the \arr\ \texttt{f$_t$}, their relationship follows the \emph{modulus relation}:
\begin{equation}\label{eq:modulus}
    \texttt{f}_t.\textbf{count}\ \%\ (\textbf{int}(\texttt{f}_s.\textbf{bytes})\ [(-\ |\ +)\ 1])\ =\ 0
\end{equation}

\subsubsection{Product}
When an atomic field contained in an array \texttt{f$_t$} has a chain of field dependencies \{\texttt{f}$_{s1}$, \texttt{f}$_{s1}$, $\ldots$, \texttt{f}$_{sk}$\},, and the product of their values equals the count of \texttt{f$_t$}, their relationship follows the \emph{product relation}:
\begin{equation}\label{eq:product}
    \texttt{f}_t.\textbf{count}\ =\ \textbf{int}(\texttt{f}_{s1}.\textbf{bytes})\ *\ \ldots\  *\ \textbf{int}(\texttt{f}_{sn}.\textbf{bytes})
\end{equation}
This relation signifies that the count of a field, potentially part of an array, is determined by combining the values of multiple fields. This is often implemented in programs using nested loops and conditional checks. For example, in the BMP file format, the total number of pixels is calculated by multiplying the values of the \texttt{height} and \texttt{width} fields. The \texttt{bmp\_parser} program~\cite{bmp-parser-program} reads the pixel array of a BMP image using nested loops: the outer loop is controlled by the \texttt{height} field and the inner loop is controlled by the \texttt{width} field. This pattern holds for many image deserialization programs.

For efficiency, product relations are derived from modulus relations. For instance, if an array \texttt{f}$_t$ exhibits modulus relations with fields \texttt{f}$_{s1}$ and \texttt{f}$_{s2}$, meaning that \texttt{f}$_t$\texttt{.\textbf{count}} \% \texttt{\textbf{int}(f}$_{s1}$\texttt{.\textbf{bytes})} $=$ $0$ and \texttt{f}$_t$\texttt{.\textbf{count}} \% \texttt{\textbf{int}(f}$_{s2}$\texttt{.\textbf{bytes})} = $0$, then \texttt{f}$_t$\texttt{.\textbf{count}} is a multiple of both values. This implies \texttt{f}$_{s1}$ and \texttt{f}$_{s2}$ act as multipliers for the count. Once identified as multipliers, the product of their values is matched against the count of \texttt{f}$_t$ to establish their product relation.

This paper shows that fields accessed within control-dependent basic blocks have semantic dependence, inferred from program structure. This links semantic dependencies to control dependence, significantly reducing the search space for identifying fields with semantic constraints, which would otherwise require exhaustive analysis. Notably, the proposed method uses exhaustive search only when no clear semantic dependence between fields exists.

%% file: implementation.tex
\section{Implementation Details}
\label{sec:implementation}
The algorithm has been implemented in a system named \byteriitwo (\textbf{R}everse \textbf{E}ngineering \textbf{S}tructure and \textbf{SEM}antics of Binary Executable). It has five independent components:
\begin{itemize}
    \item \emph{Taint-Tracker} executes a program by tainting an input it receives and producing a taint trace as a JSON file.
    \item \emph{Taint-Analyzer} produces a TIG tree in AT\&T DOT format as well as a set of frontiers from the tree as a JSON file.
    \item \emph{Structure-Extractor} constructs the structure of the input as a text file from a set of frontiers.
    \item \emph{ICDG-Extractor} produces ICDG in AT\&T DOT format from a program binary.
    \item \emph{Semantic-Analyzer} accepts the structure of input and ICDG as input, identifies the semantic dependence between fields, and then infers the semantic constraints. The recovered structure with semantic relations is written into a text file.
\end{itemize}

Taint-Tracker and Taint-Analyzer components perform the `Partition Input into Fields' step of the algorithm (Figure~\ref{fig:algorithm-steps}). Structure-Extractor performs the `Construct Structure' step of the algorithm. ICDG-Extractor and Semantic-Analyzer perform the remaining two steps of the algorithm.

We use the Taint-Tracker of \byteri~\cite{narasimha2022}, and we implemented a Taint-Analyzer that partitions input into fields. The Taint-Tracker is written in C++ using Intel PinTool~\cite{luk2005pin} and can trace the execution of x86-64 Linux binaries. The Taint-Analyzer is written in Python 3.9. We implemented the \regex algorithm in~\cite{construct_regex_2024} as the Structure-Extractor, which is written in Python 3.9. In the ICDG-Extractor, we implemented the ICDG algorithms~\cite{sinha01} as a Ghidra~\cite{ghidra-re} script, which is written in Jython 2.7. We implemented the Semantic-Analyzer in Python 3.9.

The current implementation of Taint-Tracker supports tracing file inputs to binary executables. It requires engineering efforts to support the network and other forms of input. \byteriitwo requires only one well-formed input to a binary executable to produce the taint trace, recover structure, and infer semantic relations between fields.

%% file: evaluation.tex
\section{Experiments and Evaluation}
\label{sec:evaluation}

Here we present the results of evaluating the efficacy of our algorithm, as assessed by the following research questions:

\begin{enumerate}[label=\textbf{RQ \arabic*.}]
    \item How \emph{accurate} are the recovered structures and semantic constraints between fields inferred by the algorithm?
    \item How \emph{practical} is the algorithm for recovering structures and inferring the semantic constraints between fields of the data formats of real-world applications?
\end{enumerate}

In the following sections, we present the results of experiments to investigate the above research questions.

\input{evaluation/rq-1}

\input{evaluation/rq-2}

%% file: evaluation/rq-1.tex
\subsection{RQ 1. Accuracy of Recovered Structures and Semantic Relations}
\label{eval:rq-1}

\input{evaluation/rq-1/introduction}

\input{evaluation/rq-1/subject-programs}

\input{evaluation/rq-1/evaluation}

\input{evaluation/rq-1/results}

%% file: evaluation/rq-1/introduction.tex
To evaluate the recovered structure and semantic constraints inferred by the algorithm in a controlled experiment, we prioritized the diversity of the data formats and implementations of the subject programs. We chose the synthetic subject programs from the \byterii paper~\cite{narasimha2022} for the following reasons:
\begin{enumerate*}[(i)]
    \item They have a sufficiently diverse collection of synthetic programs that cover a large variety of data formats and a variety of ways of parsing those data.
    \item Choosing the synthetic programs from the \byterii paper enables us to perform a \emph{controlled experiment} and also make a direct comparison with their results.
\end{enumerate*}

The \byterii paper evaluated the accuracy of its algorithm for recovering the structure of data formats. Although their method only recovers the structure and not the semantic relations between fields, their synthetic program set includes several programs that accept inputs with semantic relations. Therefore, we only considered the synthetic subject programs from the \byterii paper that accept input with at least one semantic relation between its fields in our evaluation.

The rest of this section describes the following:
\begin{enumerate*}[(i)]
    \item the properties of the chosen synthetic programs,
    \item the method we employed to evaluate the accuracy of the recovered structure, and
    \item the results of our evaluation.
\end{enumerate*}

%% file: evaluation/rq-1/subject-programs.tex
\subsubsection{Synthetic Subject Programs}
\label{rq-1-subject-programs}
\input{tables/program-types}

We used a subset of the synthetic subject programs from~\cite{narasimha2022} to perform a controlled experiment. The data formats of the chosen synthetic programs can be categorized into five semantic relations and two types of data encoding: ASCII and Binary.

The synthetic subject programs were written to highlight the different methods by which several real-world programs access arrays, as the structural features of programs accessing arrays differ, and array access impacts taint propagation.

The chosen programs are various combinations of encoding, semantic relations, and access methods. These programs are compiled using GCC and the resulting binaries are used as subject programs. The semantic relations, access methods, and the eight programs are described below.

The semantic relations labelled \textbf{R1} to \textbf{R7} are as follows:
\begin{enumerate*}[label=\textbf{R\arabic* --}, afterlabel=\hspace{2pt}, itemjoin=\enskip]
    \item Size,
    \item Termination,
    \item Count,
    \item Offset,
    \item Enumeration,
    \item Modulus, and
    \item Product.
\end{enumerate*}

The six methods for accessing array elements, labelled \textbf{M1} to \textbf{M6}, are as follows:

\begin{enumerate}[label=\textbf{M\arabic*}]
    \item Index an element \textit{relative to the base pointer} of the array.
    \item Index an element \textit{relative to the previous field} or record in the array.
    \item Index an element using \textit{relative virtual address} (computed after subtracting the base address).
    \item Use multiple loops to access all the elements \textit{sequentially}.
    \item Use \textit{nested loops} to access the same array.
    \item Use \textit{recursion} to access array elements.
\end{enumerate}

The properties of the chosen eight synthetic subject programs for the evaluation are summarized in Table~\ref{table:program-types}.

All the \texttt{CSV\_*} and \texttt{BMP\_CSV} programs that accept CSV data and the \texttt{HTTP} program that accept HTTP header requests in Table~\ref{table:program-types}, accept input with ASCII encodings, where the values and records are demarcated by fixed values. The \texttt{CSV} and \texttt{CSV\_Recursive\_001} programs accept inputs with the same data format of the form: \texttt{[num[,num]*\escape{n}]*}, and they access arrays in loops or recursion.

\justify{}
The three programs, \texttt{CSV\_Array}, \texttt{CSV\_Nested\_Array}, and \texttt{CSV\_Array\_Recursive}, accept CSV data of a similar form to the data format of the \texttt{sum\_csv} in Listing~\ref{lst:sum_csv}. These programs utilize \textbf{M1}, \textbf{M4}, \textbf{M5}, and \textbf{M6} access methods to process the input data. The data format encompasses a count relationship and two termination relationships, and its regular expression is of the form: \texttt{[count[,num]\{count\}\escape{n}]*}.

The \texttt{BMP\_CSV} program handles BMP images encoded in CSV format. This data format incorporates two modulus relationships, a product relationship, a count relationship, and multiple termination relationships. To process the input data, the program employs \textbf{M1}, \textbf{M4}, and \textbf{M5} access methods.

The input of the \texttt{PE} program is an array of records. Each record contains a field that has an offset relationship with a variable-length string in the data section and another field that specifies the string's size. Strings in the data section have a fixed value as a terminator. Thus, each string in the data section has three relations: offset, size, and termination.

The \texttt{PNG-2} program's input is an array of variant records. Each record contains a record-type field with an enumeration relationship with the array and multiple variable-length strings, where each string is preceded by its size field.

The input data for both the \texttt{PE} and \texttt{PNG-2} program is encoded in binary.

%% file: tables/program-types.tex
\begin{table}
	\centering
	\begin{adjustbox}{max width=1.0\linewidth}
		\begin{tabular}{lcccccccccccccccccc}
			\hline
			\multicolumn{1}{|l?}{\begin{tabular}[c]{@{}l@{}}Program \\
					Name\end{tabular}} &
			\multicolumn{7}{c?}{Semantic Relations}         &
			\multicolumn{2}{c?}{\begin{tabular}[c]{@{}c@{}}
					Data \\ encoding\end{tabular}}  &
			\multicolumn{6}{c|}{Methods of access}
			\\ \hline

			\multicolumn{1}{|l?}{}                          &
			\multicolumn{1}{l|}{R1}                         &
			\multicolumn{1}{l|}{R2}                         &
			\multicolumn{1}{l|}{R3}                         &
			\multicolumn{1}{l|}{R4}                         &
			\multicolumn{1}{l|}{R5}                         &
			\multicolumn{1}{l|}{R6}                         &
			\multicolumn{1}{l?}{R7}                         &
			\multicolumn{1}{l|}{ASCII}                      &
			\multicolumn{1}{l?}{Raw}                        &
			\multicolumn{1}{l|}{M1}                         &
			\multicolumn{1}{l|}{M2}                         &
			\multicolumn{1}{l|}{M3}                         &
			\multicolumn{1}{l|}{M4}                         &
			\multicolumn{1}{l|}{M5}                         &
			\multicolumn{1}{l|}{M6}
			\\ \hline

			\multicolumn{1}{|l?}{CSV}                       &
			\multicolumn{1}{l|}{}                           &
			\multicolumn{1}{c|}{\faCheckCircle}             &
			\multicolumn{1}{c|}{}                           &
			\multicolumn{1}{l|}{}                           &
			\multicolumn{1}{c|}{}                           &
			\multicolumn{1}{l|}{}                           &
			\multicolumn{1}{l?}{}                           &

			\multicolumn{1}{c|}{\faCheckCircle}             &
			\multicolumn{1}{l?}{}                           &

			\multicolumn{1}{l|}{}                           &
			\multicolumn{1}{c|}{\faCheckCircle}             &
			\multicolumn{1}{l|}{}                           &
			\multicolumn{1}{c|}{\faCheckCircle}             &
			\multicolumn{1}{l|}{}                           &
			\multicolumn{1}{l|}{}
			\\ \hline

			\multicolumn{1}{|l?}{CSV\_Array}                &
			\multicolumn{1}{c|}{}                           &
			\multicolumn{1}{c|}{\faCheckCircle}             &
			\multicolumn{1}{c|}{\faCheckCircle}             &
			\multicolumn{1}{l|}{}                           &
			\multicolumn{1}{l|}{}                           &
			\multicolumn{1}{l|}{}                           &
			\multicolumn{1}{l?}{}                           &

			\multicolumn{1}{c|}{\faCheckCircle}             &
			\multicolumn{1}{l?}{}                           &

			\multicolumn{1}{c|}{\faCheckCircle}             &
			\multicolumn{1}{l|}{}                           &
			\multicolumn{1}{l|}{}                           &
			\multicolumn{1}{c|}{\faCheckCircle}             &
			\multicolumn{1}{l|}{}                           &
			\multicolumn{1}{l|}{}
			\\ \hline

			\multicolumn{1}{|l?}{CSV\_Nested\_Array}        &
			\multicolumn{1}{c|}{}                           &
			\multicolumn{1}{c|}{\faCheckCircle}             &
			\multicolumn{1}{c|}{\faCheckCircle}             &
			\multicolumn{1}{l|}{}                           &
			\multicolumn{1}{l|}{}                           &
			\multicolumn{1}{l|}{}                           &
			\multicolumn{1}{l?}{}                           &

			\multicolumn{1}{c|}{\faCheckCircle}             &
			\multicolumn{1}{l?}{}                           &

			\multicolumn{1}{l|}{}                           &
			\multicolumn{1}{c|}{\faCheckCircle}             &
			\multicolumn{1}{l|}{}                           &
			\multicolumn{1}{l|}{}                           &
			\multicolumn{1}{c|}{\faCheckCircle}             &
			\multicolumn{1}{l|}{}
			\\ \hline

			\multicolumn{1}{|l?}{CSV\_Recursive\_001}       &
			\multicolumn{1}{l|}{}                           &
			\multicolumn{1}{c|}{\faCheckCircle}             &
			\multicolumn{1}{c|}{}                           &
			\multicolumn{1}{l|}{}                           &
			\multicolumn{1}{l|}{}                           &
			\multicolumn{1}{l|}{}                           &
			\multicolumn{1}{l?}{}                           &

			\multicolumn{1}{c|}{\faCheckCircle}             &
			\multicolumn{1}{l?}{}                           &

			\multicolumn{1}{l|}{}                           &
			\multicolumn{1}{l|}{}                           &
			\multicolumn{1}{l|}{}                           &
			\multicolumn{1}{l|}{}                           &
			\multicolumn{1}{l|}{}                           &
			\multicolumn{1}{c|}{\faCheckCircle}
			\\ \hline

			\multicolumn{1}{|l?}{CSV\_Array\_Recursive}     &
			\multicolumn{1}{c|}{}                           &
			\multicolumn{1}{c|}{\faCheckCircle}             &
			\multicolumn{1}{c|}{\faCheckCircle}             &
			\multicolumn{1}{l|}{}                           &
			\multicolumn{1}{l|}{}                           &
			\multicolumn{1}{l|}{}                           &
			\multicolumn{1}{l?}{}                           &

			\multicolumn{1}{c|}{\faCheckCircle}             &
			\multicolumn{1}{l?}{}                           &

			\multicolumn{1}{l|}{}                           &
			\multicolumn{1}{l|}{}                           &
			\multicolumn{1}{l|}{}                           &
			\multicolumn{1}{l|}{}                           &
			\multicolumn{1}{l|}{}                           &
			\multicolumn{1}{c|}{\faCheckCircle}
			\\ \hline

			\multicolumn{1}{|l?}{HTTP}                      &
			\multicolumn{1}{c|}{}                           &
			\multicolumn{1}{c|}{\faCheckCircle}             &
			\multicolumn{1}{c|}{}                           &
			\multicolumn{1}{l|}{}                           &
			\multicolumn{1}{l|}{}                           &
			\multicolumn{1}{l|}{}                           &
			\multicolumn{1}{l?}{}                           &

			\multicolumn{1}{c|}{\faCheckCircle}             &
			\multicolumn{1}{l?}{}                           &

			\multicolumn{1}{c|}{\faCheckCircle}             &
			\multicolumn{1}{l|}{}                           &
			\multicolumn{1}{l|}{}                           &
			\multicolumn{1}{c|}{\faCheckCircle}             &
			\multicolumn{1}{l|}{}                           &
			\multicolumn{1}{l|}{}
			\\ \hline

			\multicolumn{1}{|l?}{BMP\_CSV}                   &
			\multicolumn{1}{c|}{}                           &
			\multicolumn{1}{c|}{\faCheckCircle}             &
			\multicolumn{1}{c|}{\faCheckCircle}             &
			\multicolumn{1}{l|}{}                           &
			\multicolumn{1}{l|}{}                           &
			\multicolumn{1}{l|}{\faCheckCircle}             &
			\multicolumn{1}{l?}{\faCheckCircle}             &

			\multicolumn{1}{c|}{\faCheckCircle}             &
			\multicolumn{1}{l?}{}                           &

			\multicolumn{1}{c|}{\faCheckCircle}             &
			\multicolumn{1}{l|}{}                           &
			\multicolumn{1}{l|}{}                           &
			\multicolumn{1}{c|}{\faCheckCircle}             &
			\multicolumn{1}{l|}{\faCheckCircle}             &
			\multicolumn{1}{l|}{}
			\\ \hline

			\multicolumn{1}{|l?}{PE}                        &
			\multicolumn{1}{c|}{\faCheckCircle}             &
			\multicolumn{1}{l|}{}                           &
			\multicolumn{1}{c|}{}                           &
			\multicolumn{1}{c|}{\faCheckCircle}             &
			\multicolumn{1}{l|}{}                           &
			\multicolumn{1}{l|}{}                           &
			\multicolumn{1}{l?}{}                           &

			\multicolumn{1}{c|}{}                           &
			\multicolumn{1}{c?}{\faCheckCircle}             &

			\multicolumn{1}{c|}{\faCheckCircle}             &
			\multicolumn{1}{l|}{}                           &
			\multicolumn{1}{c|}{\faCheckCircle}             &
			\multicolumn{1}{c|}{\faCheckCircle}             &
			\multicolumn{1}{l|}{}                           &
			\multicolumn{1}{l|}{}
			\\ \hline

			\multicolumn{1}{|l?}{PNG-2}                     &
			\multicolumn{1}{c|}{\faCheckCircle}             &
			\multicolumn{1}{l|}{}                           &
			\multicolumn{1}{c|}{}                           &
			\multicolumn{1}{l|}{}                           &
			\multicolumn{1}{c|}{\faCheckCircle}             &
			\multicolumn{1}{l|}{}                           &
			\multicolumn{1}{l?}{}                           &

			\multicolumn{1}{l|}{}                           &
			\multicolumn{1}{c?}{\faCheckCircle}             &

			\multicolumn{1}{l|}{}                           &
			\multicolumn{1}{c|}{\faCheckCircle}             &
			\multicolumn{1}{l|}{}                           &
			\multicolumn{1}{c|}{\faCheckCircle}             &
			\multicolumn{1}{l|}{}                           &
			\multicolumn{1}{l|}{}
			\\ \hline
		\end{tabular}
	\end{adjustbox}
	\captionsetup{justification=centering, font=small}
	\caption{Details of the synthetic programs~\cite{narasimha2022}.}
	\label{table:program-types}
\end{table}

%% file: evaluation/rq-1/evaluation.tex
\subsubsection{Method for Evaluation}
\label{rq-1-evaluation-method}

We evaluate the accuracy of the recovered structures and semantic relations between fields using two methods:

\paragraph{Manual Inspection.} We compare the recovered structure with an expected structure.

\paragraph{Validity of Generated Data.} We generate 1,000 new random data using each inferred structure and feed each generated data to the original subject program. If the program accepts the generated data without raising any errors, we consider the generated data to be valid. The accuracy of a recovered structure is measured with the acceptance ratio, that is the percentage of data accepted by the original subject program.

We also compare the acceptance ratio of \byteriitwo to the acceptance ratio of the \byterii system for the same set of synthetic programs using the inferred RSMs for generating data and feeding data to the original synthetic subject programs.

%% file: evaluation/rq-1/results.tex
\subsubsection{Results}
\label{rq-1-results}
\input{tables/acceptance_ratio}

\paragraph{Manual Inspection.}
Per the manual review, the structures inferred by \byteriitwo for all the synthetic programs matched the corresponding actual data format. The algorithm correctly identifies the syntactic components of the structure, such as fixed-length fields, variable-length fields, arrays of atomic fields, arrays of records, arrays in nested arrays, and arrays with variant records. Furthermore, the algorithm correctly identifies semantic relations like size, count, offset, and enumeration. Thus, the inferred structure correctly reflects the actual syntax and semantics of the data.

\paragraph{Validity of Generated Data.}
Table~\ref{table:set-acceptance-rate} presents the acceptance rates for the 1,000 generated input data of \byterii~\cite{narasimha2022} and \byteritwo.

\byteriitwo generated acceptable data for all subject programs, while \byterii generated acceptable data for programs that accept inputs with termination relations.

Specifically, the data generated using \byteritwo's inferred structures achieves a 100\% acceptance rate for all subject programs. This is because \byteriitwo infers the semantic relations between fields that are required to generate valid data acceptable to the original subject program.

\byterii does not infer any semantic relations, its acceptance rate is abysmally poor for most subject programs. The acceptance rates are non-zero because randomly generated data rarely meets semantic constraints.

\byterii generated valid data for subject programs that accept input with only termination relations, such as \texttt{HTTP}, \texttt{CSV}, and \texttt{CSV\_Recursive\_001} programs. In the inferred RSMs of these data formats, the termination field immediately follows the variable-length field, and there are no other outgoing edges from the variable-length field to other fields in the state machine. This means that the generated data using the inferred RSMs is guaranteed to satisfy the constraint that the value that follows the generated value of a variable-length field matches the terminator value.

%% file: tables/acceptance_ratio.tex
\begin{table}
	\centering
	\begin{adjustbox}{max width=0.45\linewidth}
		\begin{tabular}{|l|r|r|}
			\hline
			\multicolumn{1}{|l|}{} &
			\multicolumn{2}{c|}{Acceptance Ratio}
			\\ \hline
			Program                & \byterii & \byteritwo \\ \hline \hline
			CSV                    & 100\%    & 100\%      \\ \hline
			CSV\_Array             & 5.3\%    & 100\%      \\ \hline
			CSV\_Nested\_Array     & 5.0\%    & 100\%      \\ \hline
			CSV\_Recursive\_001    & 100\%    & 100\%      \\ \hline
			CSV\_Array\_Recursive  & 2.6\%    & 100\%      \\ \hline
			HTTP                   & 100\%    & 100\%      \\ \hline
			BMP\_CSV               & 1.6\%    & 100\%      \\ \hline
			PE                     & 1.2\%    & 100\%      \\ \hline
			PNG-2                  & 3.4\%    & 100\%      \\ \hline
		\end{tabular}
	\end{adjustbox}
	\captionsetup{justification=centering, font=small}
	\caption{Inputs generated by inferred RSMs that were accepted by the program (1,000 inputs each)}
	\label{table:set-acceptance-rate}
\end{table}

%% file: evaluation/rq-2.tex
\subsection{RQ 2. Algorithm's Practicability}
\label{eval:rq-2}

\input{evaluation/rq-2/introduction}

\input{evaluation/rq-2/results}

\input{evaluation/rq-2/use-case}

%% file: evaluation/rq-2/introduction.tex
To assess the practicality of our algorithm and determine whether it is a viable solution for real-world programs, we perform three evaluations:

\begin{enumerate}
    \item \textbf{Time Performance.} Measure the time it takes to apply \byteriitwo to synthetic subject programs and selected real-world programs.
    \item \textbf{Validity of Generated Data}. Generate data using the recovered structures and verify that the generated data is accepted by the original subject program.
    \item \textbf{Comparison with Ground Truth}. Compare the \byteriitwo inferred structure of the input of a real-world program to the actual structure in the data format.
\end{enumerate}

\subsubsection{Real-world subject programs.}
We selected the following command line utility tools, compiled using GCC as \texttt{x86-64} Linux binaries:
\begin{enumerate*}[(a), afterlabel=\hspace{2pt}, itemjoin=\enskip]
    \item \texttt{bmp-parser} -  validates BMP files,
    \item \texttt{pngcheck} - verifies the integrity of PNG files, and
    \item \texttt{unzip} - displays the metadata information of ZIP files.
\end{enumerate*}

%% file: evaluation/rq-2/results.tex
\subsubsection{Time Performance}
\label{rq-2-time-perf}
\input{tables/time_performance_results}

Table~\ref{table:time-performance-results} shows the performance results for \byteriitwo on both the synthetic and real-world subject programs. The input sizes range from 61 bytes to 3.1 MB. The subject programs were run on a 2 GHz AMD Ryzen 7 5850U machine running Ubuntu 20.04 LTS. The table shows the execution time broken down into the four stages of the \byteriitwo algorithm:
\begin{enumerate*}[(i), afterlabel=\hspace{2pt}, itemjoin=\enskip]
    \item Partition Input into Fields,
    \item Compute Inter-Procedural Control Dependence Graph (ICDG),
    \item Construct Structure, and
    \item Infer Semantic Relations (combines Identify Semantic Dependence between Fields and Infer Semantic Relations).
\end{enumerate*}

The native execution time of all the subject programs is a fraction of a second, so it is not shown in the table. With the data shown in the table, the Partition Input into Fields stage takes several orders of magnitude longer than the other stages. The major component within the Partition Input into Field is the Taint Tracker. The time it takes to perform taint tracking increases with the size of the binary or the total number of instructions executed. Therefore, the Partition Input into Fields stage dominates the cost of the \byteriitwo algorithm and controls its practical application.

\subsubsection{Validity of generated data}
\label{rq-2-programmatic-eval}

For this experiment, we generated 1,000 new input data using the recovered structure and fed each generated data to the original program. The acceptance ratios of the chosen real-world programs are as follows:
\begin{enumerate*}[(i), afterlabel=\hspace{2pt}, itemjoin=\enskip]
    \item \texttt{bmp\_parser}: 100\%,
    \item \texttt{unzip}: 100\%, and
    \item \texttt{pngcheck}: 0\%.
\end{enumerate*}

\byteriitwo generated data for the \texttt{bmp\_parser} and \texttt{unzip} programs was acceptable, while all the data generated for the \texttt{pngcheck} were not acceptable. \byteriitwo infers the semantic relations between fields that are required to generate valid data acceptable to the \texttt{bmp\_parser} and \texttt{unzip} programs. However, for the \texttt{pngcheck} program, although \byteriitwo infers the required semantic relations between fields, it over-generalized the values of a control field, resulting in invalid generated data.

The \texttt{bmp\_parser} program processes all of the input fields and uses the two semantic relations, offset and product relations, to process the pixel array in the data section. \byteriitwo recovered all the syntactic components of the BMP structure and the two semantic relations. The recovered structure of BMP can be found in Appendix~\ref{appendix:use-case-bmp}: Listing~\ref{lst:bmp_struct}.

The \texttt{unzip -Z} or \texttt{zipinfo} program processes the metadata or directory entries of each file within a ZIP file. Within each ZIP directory, there is a semantic relationship between two fields: a size field (\texttt{fileNameLength}) that specifies the length of a variable-length field (\texttt{fileName}) in which the name of the file is stored. The ZIP file format also includes a trailer section with fields that specify the total number of files, the size of the data section, and the start offset of the first file directory within the ZIP file. However, the \texttt{zipinfo} program searches for the directory signature to locate the first file directory. As a result, \byteriitwo was not able to infer any of these three relations. Therefore, any input data that just satisfies the size relation of the \texttt{fileName} field was sufficient for the \texttt{zipinfo} program to be acceptable. The recovered structure of ZIP can be found in Appendix~\ref{appendix:use-case-zip}: Listing~\ref{lst:zip_struct}.

The PNG format contains a signature field followed by PNG chunks, which is an array of variant records. Each PNG chunk contains the following fields:
\begin{enumerate*}[(i), afterlabel=\hspace{2pt}, itemjoin=\enskip]
    \item A 4-byte \texttt{length} field that specifies the size of the data field,
    \item A 4-byte \texttt{type} field that specifies the record type in the data,
    \item A variable-length \texttt{data} field that contains the actual data or a record, and
    \item A 4-byte \texttt{CRC} field that specifies the cyclic redundancy check value of the data in the chunk.
\end{enumerate*}
\byteriitwo identifies all the syntactic components of each PNG chunk, including the structures within the data field. It also identifies that chunks are variant records, infers the size relationship between the \texttt{length} field and \texttt{data} field of most chunks for which the program validates their size, and infers that \texttt{type} field in each chunk specifies the type of record.

The first eight bytes (\texttt{89 50 4E 47 0D 0A 1A 0A}) of a PNG file are the signature bytes of the file format. The \texttt{pngcheck} program while processing the file signature of a PNG image, verifies ASCII printable bytes (\texttt{50 4E 47} or ASCII: \texttt{P N G}) of the signature bytes in a loop using the same set of instructions. Therefore, the Taint-Analyzer treats them as an array of a 1-byte field and over-generalizes their values to create a token, \texttt{UPPER}, a set of upper case ASCII alphabets. The data generated using the recovered structure of PNG contains incorrect signature bytes. Such input data are immediately discarded by the program. Marrying \byteriitwo with fuzzers that identify signature bytes, such as, ~\cite{li_steelix_2017,afl-analyze,cho_intriguer_2019}, may prevent over-generalizing those bytes. The recovered structure of PNG can be found in Appendix~\ref{appendix:use-case-png}: Listing~\ref{lst:png_struct}.

%% file: tables/time_performance_results.tex
\begin{table}
	\centering
	\begin{adjustbox}{max width=1.0\linewidth}
		\begin{filecontents*}{tables/time_performance_results.csv}
name,instructions,binsize,inputsize,partition-input,structure-time,icdg-time,semantics-time
CSV,238536,17184,61,1.47,0.003,0.154,0.001
CSV\_Array,236471,17184,75,1.588,0.006,0.162,0.001
CSV\_Nested\_Array,275463,17272,103,1.635,0.004,0.202,0.001
CSV\_Array\_Recursive,256230,17168,111,1.576,0.009,0.154,0.001
CSV\_Recursive\_001,247962,17448,61,1.513,0.004,0.159,0.001
HTTP,260887,17472,511,1.778,0.021,0.232,0.001
PE,251756,22792,136,1.535,0.003,0.333,0.001
PNG-2,447112,23224,328,2.695,0.036,0.381,0.015
bmp-parser,333392,17232,86,1.644,0.022,0.258,0.002
pngcheck (-vt),318927,129600,1049,2.966,0.011,9.112,0.002
unzip (-Z),638056,210600,3235684,11.72,0.297,5.009,0.261
gifsicle (-I),2434623,906256,8703,19.119,0.829,6.875,0.066
\end{filecontents*}
\pgfplotstabletypeset[
			every nth row={8}{before row=\hline\hline},
			columns={
					name,binsize,inputsize,partition-input,icdg-time,structure-time,semantics-time
				},
			column type/.add={}{|},
			columns/binsize/.style={
					column name={\stacklabel{Binary\\Size\\(bytes)}},
					column type=r|,
					int detect,
				},
			columns/inputsize/.style={
					column name={\stacklabel{Input\\size\\(bytes)}},
					column type=r?,
					int detect,
				},
			columns/name/.style={
					column type=|l|,
					column name={Program Name},
					string type,
				},
			columns/instructions/.style={
					column name={\stacklabel{Instructions\\Executed\\(\#)}},
					column type=r|,
					int detect,
				},
			columns/partition-input/.style={
					column name={\stacklabel{Partition\\Input into\\Fields\\(seconds)}},
					fixed,
					fixed zerofill,
					column type=r?,
					precision=3,
				},
			columns/structure-time/.style={
					column name={\stacklabel{Construct\\Structure\\Time\\(seconds)}},
					fixed,
					fixed zerofill,
					column type=r?,
					precision=3,
				},
			columns/icdg-time/.style={
					column name={\stacklabel{Compute\\ICDG\\Time\\(seconds)}},
					fixed,
					fixed zerofill,
					column type=r?,
					precision=3,
				},
			columns/semantics-time/.style={
					column name={\stacklabel{Infer\\Semantic\\Relations\\Time\\(seconds)}},
					fixed,
					fixed zerofill,
					column type=r|,
					precision=3,
				},
		]{tables/time_performance_results.csv}
	\end{adjustbox}
	\captionsetup{justification=centering, font=small}
	\caption{Performance measurement of \byteriitwo for synthetic and real-world programs}
	\label{table:time-performance-results}
\end{table}

%% file: evaluation/rq-2/use-case.tex
\subsubsection{Comparison with Ground Truth}
\label{rq-2:comparison-with-ground-truth}

\paragraph{\textbf{010 Editor as Ground Truth}}

We use the input structure knowledge provided by 010 Editor~\cite{010editor} as the ground truth. 010 Editor is a commercial off-the-shelf (COTS) application that can parse and display the structure of several well-known data formats. The parsers or data format templates in 010 Editor are written by data format experts and verified by the community members. Therefore, the structures of input files of several well-known data formats parsed by 010 Editor are used as the ground truth for the evaluation of \byteriitwo recovered structures with semantics.

\paragraph{Use Case: BMP image}

We compared the structure of BMP recovered by \byteriitwo with the structure of BMP parsed by the ground truth. \byteriitwo identifies all the syntactic components of the BMP structure as well as the two semantic relations. A detailed comparison of the structure of BMP inferred by \byteriitwo and the ground truth can be found in Appendix~\ref{appendix:use-case-bmp}.

%% file: related-works.tex
\section{Related Works}
\label{sec:related-works}

\input{related-works/introduction}

%% file: related-works/introduction.tex
\input{tables/cmp-structure}

We divide the related work into four groups: (a) network protocol reverse engineering, (b) format-aware fuzzing, (c) learning context-free grammar, and (d) recovering state machine representation of input.

We compare our work with the other works in the context of their ability to recover the syntactic structure of input and infer semantic relations between fields of input.

The syntactic components of structures are:
\begin{enumerate*}[label=\textbf{S\arabic* --}, afterlabel=\hspace{2pt}, itemjoin=\enskip]
    \item Values;
    \item Variable length strings;
    \item Variable length ASCII-encoded numbers;
    \item Arrays of atomic fields;
    \item Arrays of composite fields;
    \item Arrays with variant records;
    \item Arbitrarily nested structures;
    \item Context-free grammar; and
    \item Nested structure (state machine).
\end{enumerate*}

The semantic relations between fields are:
\begin{enumerate*}[label=\textbf{R\arabic* -- }, afterlabel=\hspace{2pt}, itemjoin=\enskip]
    \item Value determining variant record type;
    \item Terminator for a variable length field;
    \item Counter of a variable length field;
    \item Terminator for array records;
    \item Counter of an array;
    \item Field determining the offset of another field;
    \item Checksum of a data chunk; and
    \item Value from an enumerated set.
\end{enumerate*}

The thirteen related research methods that we compared against \byteriitwo are summarized in Table~\ref{table:related-methods-comparison}.

We say that a research method identifies a syntactic component or semantic relationship \textit{completely}, indicated by \OK~in Table~\ref{table:related-methods-comparison}, when it recovers under all the constraints. For instance, a method completely recovers syntactic component type \textbf{S5} when it satisfies the following constraints:
\begin{enumerate*}[label=\textbf{S5.\alph* -- }, afterlabel=\hspace{2pt}, itemjoin=\enskip]
    \item Identifies the sequence of fields that form an array.
    \item Recovers arrays of records with arbitrary levels of nesting.
    \item Recovers regardless of the method of access by the subject programs (\textbf{M1} to \textbf{M6} specified in Section~\ref{rq-1-subject-programs}).
    \item Recovers binary and ASCII encoded arrays of records.
\end{enumerate*}

In Table~\ref{table:related-methods-comparison}, apart from \byteriitwo, only two other methods, Autogram~\cite{hoschele2017mining} and Mimid~\cite{gopinath_2020}, satisfy all the above constraints for recovering syntactic component type \textbf{S5}.

Similarly, a method completely infers the semantic relationship \textbf{R3} when it satisfies the following constraints:
\begin{enumerate*}[label=\textbf{R3.\alph* -- }, afterlabel=\hspace{2pt}, itemjoin=\enskip]
    \item Identifies the source field that specifies the size of a variable-length field.
    \item Identifies the target field, which is a variable-length string, for which the size field specifies the number of bytes.
\end{enumerate*}

In Table~\ref{table:related-methods-comparison}, apart from \byteriitwo, only two other methods, Autogram~\cite{hoschele2017mining} and Mimid~\cite{gopinath_2020}, satisfy all the above constraints for recovering syntactic component type \textbf{S5}.

We say that a research method identifies a syntactic component or semantic relationship \textit{partially}, indicated by \PARTIAL~in Table~\ref{table:related-methods-comparison}, when it satisfies only a few constraints. For instance, methods like PolyGlot~\cite{caballero_polyglot_2007}, Wondracek \textit{et al.}~\cite{wondracek_2008}, Tupni~\cite{cui_tupni_2008}, \textsc{ProFuzzer}~\cite{you_profuzzer_2019}, \textsc{WEIZZ}~\cite{fioraldi_weizz_2020}, and \textsc{AIFORE}~\cite{shi_aifore_2023} identifies the semantic type as count (\textbf{R5}) of the source fields. However, these methods do not identify the target fields of count relations. The reason for this is these methods do not identify a set of values or fields that are part of the same repeating structure, array (\textbf{S5}).

Autogram~\cite{gopinath_2020}, Mimid~\cite{hoschele2017mining}, \byteri~\cite{narasimha2022}, and \byteriitwo identify array structures in the input. Of these, only \byteriitwo identifies the count relations as it identifies a count field as the source field and an array as its target field of the relationship, regardless of the access method used by the subject programs or the data encodings of inputs.

\byterii presents the recovered structure as a state machine, which requires further analysis to identify any types of arrays, and it does not infer any semantic constraints. Autogram~\cite{hoschele2017mining} and Mimid~\cite{gopinath_2020} recover arrays of atomic fields and arrays of records, but they cannot recover arrays with variant records. This is because these methods construct the structure of the context-free grammar based on the structure of the subject program, which enables them to identify syntactic constraints but not semantic constraints. Arrays with variant records have semantic constraints, as the type of a record in the array is determined by a field within the record. Therefore, these methods are limited to recovering structures with no semantic constraints. \byteriitwo does not learn context-free grammar, but it recovers a structure equivalent to the state machine representation of the input, and recovers arrays with variant records, as well as identifies the record type field.

%% file: tables/cmp-structure.tex
\begin{table}
    \centering
    \begin{adjustbox}{max width=1.0\linewidth}
        \begin{tabular}{|c|c|c|c|c|c|c|c|c|c|c?c|c|c|c|c|c|c|c|}
            \hline
            Scope                                         &
            Method                                        &
            \multicolumn{9}{c?}{Syntactic Components}    &
            \multicolumn{8}{c|}{Semantic Relations}
            \\ \hline
            \rot{}                                        &
            \rot{}                                        &
            S1                                            &
            S2                                            &
            S3                                            &
            S4                                            &
            S5                                            &
            S6                                            &
            S7                                            &
            S8                                            &
            S9                                            &
            R1                                            &
            R2                                            &
            R3                                            &
            R4                                            &
            R5                                            &
            R6                                            &
            R7                                            &
            R8
            \\ \hline
            \hline
            \multirow{4}{*}{
                \begin{tabular}[c]{@{}c@{}}
                    Protocol
                \end{tabular}
            }                                             &
            PolyGlot~\cite{caballero_polyglot_2007}       &
            \OK                                           &
            \OK                                           &
                                                          &
            \PARTIAL                                      &
            \PARTIAL                                      &
                                                          &
                                                          &
                                                          &
                                                          &
                                                          &
            \PARTIAL                                      &
            \PARTIAL                                      &
                                                          &
            \PARTIAL                                      &
                                                          &
                                                          &
            \PARTIAL
            \\ \cline{2-19}
            \rot{}                                        &
            Wondracek \emph{et al.}~\cite{wondracek_2008} &
            \OK                                           &
            \OK                                           &
                                                          &
            \PARTIAL                                      &
            \PARTIAL                                      &
                                                          &
                                                          &
                                                          &
                                                          &
                                                          &
            \PARTIAL                                      &
            \PARTIAL                                      &
                                                          &
            \PARTIAL                                      &
                                                          &
                                                          &

            \\ \cline{2-19}
            \rot{}                                        &
            AutoFormat~\cite{lin2010}                     &
            \OK                                           &
            \OK                                           &
            \PARTIAL                                      &
            \PARTIAL                                      &
            \PARTIAL                                      &
                                                          &
                                                          &
                                                          &
                                                          &
                                                          &
                                                          &
                                                          &
                                                          &
                                                          &
                                                          &
                                                          &

            \\ \cline{2-19}                                              &
            Tupni~\cite{cui_tupni_2008}                   &
            \OK                                           &
            \OK                                           &
                                                          &
            \PARTIAL                                      &
            \PARTIAL                                      &
            \PARTIAL                                      &
                                                          &
                                                          &
                                                          &
                                                          &
                                                          &
            \PARTIAL                                      &
            \PARTIAL                                      &
            \PARTIAL                                      &
                                                          &
            \PARTIAL                                      &

            \\ \hline \hline
            \multirow{7}{*}{
                \begin{tabular}[c]{@{}c@{}}
                    Format \\ Fuzzing
                \end{tabular}
            }
                                                          &
            Intriguer~\cite{cho_intriguer_2019}           &
            \OK                                           &
            \OK                                           &
                                                          &
                                                          &
                                                          &
                                                          &
                                                          &
                                                          &
                                                          &
                                                          &
                                                          &
                                                          &
                                                          &
                                                          &
                                                          &
            \PARTIAL                                      &
            \PARTIAL
            \\ \cline{2-19}
                                                          &
            AFL-Analyze~\cite{afl-analyze}                &
            \OK                                           &
            \OK                                           &
            \PARTIAL                                      &
                                                          &
                                                          &
                                                          &
                                                          &
                                                          &
                                                          &
                                                          &
                                                          &
            \PARTIAL                                      &
                                                          &
                                                          &
                                                          &
            \PARTIAL                                      &
            \PARTIAL
            \\ \cline{2-19}
                                                          &
            TIFF-Fuzz~\cite{jain_tiff_2018}               &
            \OK                                           &
            \OK                                           &
            \PARTIAL                                      &
            \PARTIAL                                      &
            \PARTIAL                                      &
                                                          &
                                                          &
                                                          &
                                                          &
                                                          &
                                                          &
                                                          &
                                                          &
                                                          &
                                                          &
                                                          &

            \\ \cline{2-19}
                                                          &
            \textsc{ProFuzzer}~\cite{you_profuzzer_2019}  &
            \OK                                           &
            \OK                                           &
                                                          &
                                                          &
                                                          &
                                                          &
                                                          &
                                                          &
                                                          &
                                                          &
                                                          &
            \PARTIAL                                      &
                                                          &
            \PARTIAL                                      &
            \PARTIAL                                      &
                                                          &
            \PARTIAL
            \\ \cline{2-19}
                                                          &
            \textsc{WEIZZ}~\cite{fioraldi_weizz_2020}     &
            \OK                                           &
            \OK                                           &
                                                          &
                                                          &
                                                          &
                                                          &
                                                          &
                                                          &
                                                          &
                                                          &
                                                          &
            \PARTIAL                                      &
                                                          &
                                                          &
                                                          &
            \PARTIAL                                      &
            \PARTIAL
            \\ \cline{2-19}
                                                          &
            \textsc{AIFORE}~\cite{shi_aifore_2023}        &
            \OK                                           &
            \OK                                           &
                                                          &
                                                          &
                                                          &
                                                          &
                                                          &
                                                          &
                                                          &
                                                          &
                                                          &
            \PARTIAL                                      &
                                                          &
            \PARTIAL                                      &
            \PARTIAL                                      &
            \PARTIAL                                      &
            \PARTIAL
            \\ \hline \hline
            \multirow{2}{*}{
                \begin{tabular}[c]{@{}c@{}}
                    Context-Free \\ Grammar
                \end{tabular}
            }                                             &
            Autogram~\cite{hoschele2017mining}            &
            \OK                                           &
            \OK                                           &
            \OK                                           &
            \OK                                           &
            \OK                                           &
                                                          &
                                                          &
            \OK                                           &
                                                          &
                                                          &
            \OK                                           &
                                                          &
            \OK                                           &
                                                          &
                                                          &
                                                          &
            \\ \cline{2-19}
                                                          &
            Mimid~\cite{gopinath_2020}                    &
            \OK                                           &
            \OK                                           &
            \OK                                           &
            \OK                                           &
            \OK                                           &
                                                          &
                                                          &
            \OK                                           &
                                                          &
                                                          &
            \OK                                           &
                                                          &
            \OK                                           &
                                                          &
                                                          &
                                                          &

            \\ \hline \hline
            \multirow{2}{*}{
                \begin{tabular}[c]{@{}c@{}}
                    State Machine
                \end{tabular}
            }
                                                          &
            \byteri~\cite{narasimha2022}                  &
            \OK                                           &
            \OK                                           &
            \OK                                           &
            \PARTIAL                                      &
            \PARTIAL                                      &
            \PARTIAL                                      &
            \PARTIAL                                      &
                                                          &
            \OK
                                                          &
                                                          &
            \PARTIAL                                      &
                                                          &
                                                          &
                                                          &
                                                          &
                                                          &
            \PARTIAL
            \\ \cline{2-19}
                                                          &
            \byteritwo                                    &
            \OK                                           &
            \OK                                           &
            \OK                                           &
            \OK                                           &
            \OK                                           &
            \OK                                           &
            \OK                                           &
                                                          &
            \OK                                           &
            \OK                                           &
            \OK                                           &
            \OK                                           &
            \OK                                           &
            \OK                                           &
            \OK                                           &
                                                          &
            \PARTIAL
            \\ \hline
        \end{tabular}
    \end{adjustbox}
    \captionsetup{justification=centering, font=small, singlelinecheck=off}
    \caption[]{Comparison of methods in recovering the syntactic components and the semantic relations.\\
        \OK\ -- Full recovery, \PARTIAL\ -- Partial recovery
    }
    \label{table:related-methods-comparison}
\end{table}

%% file: limitations.tex
\section{limitations}
\label{sec:limitations}

\byteriitwo employs \texttt{topmost\_reps} method to extract frontiers from the TIG tree, prioritizing nodes with repeating SIs. However, this approach can lead to overgeneralized or undergeneralized grammars. Addressing these limitations may involve exploring alternative frontier selection strategies, generating grammar for multiple frontiers, and employing tests to select the optimal one, similar to Grammar fuzzing~\cite{sochor2023fuzzing,bastani2017synthesizing,bendrissou2022synthesizing,mathis2019parser}.

Our method leverages taint analysis and semantic dependence to identify a limited set of semantic relations. However, numerous other relations, such as checksums, remain uncaptured. Given the simplicity of our current approach, further research is necessary to explore more complex relationships by combining our method with other analysis techniques.

%% file: impact.tex
\section{Impact and Applications}
\label{sec:Impact}
This section briefly outlines the potential impact of this research.

The primary result of \byteritwo--- discovery of the structure of an input of a program using taint-tracking---together with the intermediate data--taint trace and interprocedural control dependence--provides the following information about a program:
\begin{description}
    \item [Control Data:] The components of the input that are used in the conditional branch instructions executed by the input.
    \item [Memory Map:] The segments of memory---stack, heap, and global---that a component of the input modifies.
    \item [Confused Integers:] The components of the input that are used in instructions with different operand sizes (8-bit, 16-bit, 32-bit, 64-bit) and a mix of signed and unsigned arithmetic instructions.
    \item [Modules:] The program instructions that use a particular component of the input.
\end{description}

The above information can improve existing and make possible new capabilities as follows:

\begin{description}
    \item [Fuzzing:] \byteritwo\ provides Control Data components of both valid and invalid inputs. A fuzzer may force a program execution along an untraversed path by fuzzing only the Control Data used in conditional branch instructions with an unexplored branch. When a branch controls a loop that processes arrays in the input, the fuzzer may automatically construct arrays of different sizes to exercise different loop boundary conditions.
    \item [Vulnerability Analysis:] \byteritwo\ provides a Memory Map of all data that affect changes on the stack and heap. Data components that are either variable lengths or arrays are candidates to fuzz to explore memory corruption vulnerabilities. Similarly, data components that are Confused Integers are good candidates to fuzz for integer overflow.
    \item [Reverse Engineering:] \byteritwo\ identifies Modules for each component of a program input. A reverse engineer may perform differential analysis---analyze the differences in the modules of two slightly different inputs---to locate instructions that implement a particular functionality.
\end{description}

%% file: conclusion.tex
\section{Conclusions and Future Work}
\label{sec:conclusion}
We presented an algorithm to recover the structure and infer the semantic constraints between fields of input acceptable by a binary program. The algorithm transforms the input, which is a sequence of bytes, into a sequence of fields and constructs the structure of the input. It uses semantic dependence between the statements of a program to infer the semantic relations between fields of input of that program. It constructs a C/C++-like structure of input to represent the syntactic components and semantic relations. We evaluated the algorithm on a set of synthetic and real-world programs and found that it was able to correctly recover the syntactical components of the structure, such as atomic field, arrays of atomic fields, arrays of records, and arrays with variant records, as well as infer the semantic relations such as size, count, offset, termination, and enumeration relations.

Our algorithm shows promise for generating new data using the inferred structures that are acceptable to the original subject program. This could be useful for a variety of tasks, including fuzzing or manual reverse engineering. Previous works~\cite{pham_smart_2019, aschermann_nautilus_2019, wang_superion_2019} have demonstrated that the availability of the input grammar can improve fuzzing. By using our algorithm to generate inputs that satisfy the semantic constraints of the program, we can improve coverage-guided fuzzing.

In future work, we plan to integrate \byteriitwo with a fuzzer to develop a more effective and efficient fuzzing tool. We believe that such a tool could be used for vulnerability hunting in a wide range of binary programs. Another possible future direction is to recover the structure of data formats defined using context-free grammar.

%% file: appendix.tex
\input{appendix/csv_array_solution}

\input{appendix/bmp-use-case}

\input{appendix/zip-use-case}

\input{appendix/png-use-case}

%% file: appendix/csv_array_solution.tex
\section{Running Example}
\label{appendix:running-example}

\begin{figure}
    \centering
    \captionsetup{justification=centering, font=small}
\begin{lstlisting}[
        style=pystyle,
        caption={Example program, \texttt{sum\_csv()}},
        label={lst:sum_csv},
    ]
fun sum_csv(input_string):
  for line in read_line(input_string) do
    counter = read_int(line, ','))
    int *values = malloc(size(int) * counter)
    int sum = 0
    for i in range(counter) do
      values[i++] = read_int(line, ',')
    for i in range(counter) do
      sum += values[i]
    print sum
\end{lstlisting}
    \Description{}{}
\end{figure}

\begin{figure}
    \centering
    \captionsetup{justification=centering, font=small}
\begin{lstlisting}[
        style=structstyle,
        caption={Recovered structure of input `4,3,2,5,8$\escape{n}$' to \texttt{sum\_csv()} program},
        label={lst:sum_csv_struct},
    ]
record S0 {
  atomic F0 (1) = <DIGIT>;
  atomic F1 (1) = <COMMA>;
  array+ A0 {
    record S1 {
      atomic F2 (1) = <DIGIT>
        WHERE F2.terminator = F3.bytes;
      atomic F3 (1) = <COMMA>;
    };
  }
    WHERE A0.count =  int(F0.bytes) - 1;
  record S2 {
    atomic F2 (1) = <DIGIT>
      WHERE F2.terminator = F4.bytes;
    atomic F4 (1) = <LINEFEED>;
  };
};
\end{lstlisting}
    \Description{}{}
\end{figure}

To illustrate the idea, the program in Listing~\ref{lst:sum_csv} when executed with the input ``4,3,2,5,8$\escape{n}$'', produces a following usage tuples:

\begin{itemize}
    \item ``4'': \{$I2$, $I3$, $I4$, $I6$, $I8$\}
    \item ``,'': \{$I2$, $I3$\}
    \item ``3'': \{$I2$, $I7$, $I9$, $I10$\}
    \item ``,'': \{$I2$, $I7$\}
    \item ``2'': \{$I2$, $I7$, $I9$, $I10$\}
    \item ``,'': \{$I2$, $I7$\}
    \item ``5'': \{$I2$, $I7$, $I9$, $I10$\}
    \item ``,'': \{$I2$, $I7$\}
    \item ``8'': \{$I2$, $I7$, $I9$, $I10$\}
    \item ``$\escape{n}$'': \{$I2$\}
\end{itemize}

where ``x'': \{$Ii$,$\ldots$,$In$\} means the byte ``x'' is used at program statements $Ii$,$\ldots$,$In$ during their respective invocations.

\subsection{Partition input into fields.}

The following are the values and fields:
\begin{itemize}
    \item \textbf{Values.} Each byte of the input forms a value. Therefore, there are 10 values for the given input.

    \item \textbf{Fields.} The following are fields and their values and the set of statements where those values are used:

          \begin{itemize}
              \item \texttt{F0} - ``4''; used at \{$I2$, $I3$, $I4$, $I6$, $I8$\}.

              \item \texttt{F1} - ``,''; used at \{$I2$, $I3$\}.

              \item \texttt{F2} - ``3'', ``2'', ``5'', ``8''; used at \{$I2$, $I7$, $I9$, $I10$\}.

              \item \texttt{F3} - ``,''; used at \{$I2$, $I7$\}.

              \item \texttt{F4} - ``$\escape{n}$''; used at \{$I2$\}.
          \end{itemize}
\end{itemize}

The input byte sequence ``4,3,2,5,8$\escape{n}$'' translates to a sequence of values: $\langle$ ``4'', ``,'', ``3'', ``,'', ``2'', ``,'', ``5'', ``,'', ``8'', ``$\escape{n}$'' $\rangle$, and further translates to a sequence of fields:
$\langle$ \texttt{F0}, \texttt{F1}, \texttt{F2}, \texttt{F3}, \texttt{F2}, \texttt{F3}, \texttt{F2}, \texttt{F3}, \texttt{F2}, \texttt{F4} $\rangle$.

\subsection{Structure of input}

Translate the sequence of fields to an element of \texttt{Structure}. The sequence of fields $\langle$ \texttt{F0}, \texttt{F1}, \texttt{F2}, \texttt{F3}, \texttt{F2}, \texttt{F3}, \texttt{F2}, \texttt{F3}, \texttt{F2}, \texttt{F4} $\rangle$ becomes the structure in Listing~\ref{lst:sum_csv_struct}.

\subsection{Fields with semantic dependence.}
\subparagraph{\textbf{Statements with control dependence.}}
\begin{itemize}
    \item $I3$--$I6$, $I8$, and $I10$ are control-dependent on $I2$ ($I2$\ $\delta^c$\ \{$I3$--$I6$, $I8$, $I10$\}).

    \item $I7$ is control dependent on $I6$ ($I6$\ $\delta^c$\ $I7$).

    \item $I9$ is control dependent on $I8$ ($I8$\ $\delta^c$\ $I9$).
\end{itemize}

\subparagraph{\textbf{Fields used at control-dependent statements.}}
\begin{itemize}
    \item \texttt{\textbf{atomic}} fields \texttt{F2} and \texttt{F3} are used at the statement $I7$ and \texttt{\textbf{atomic}} field \texttt{F2} is used at statement $I6$, thus we say that \texttt{F2} and \texttt{F3} are semantically dependent on \texttt{F0} (\texttt{F0}\ $\delta^s$\ \{\texttt{F2}, \texttt{F3}\}).

          \begin{itemize}
              \item Since the \texttt{\textbf{record}} \texttt{S1} is composed of \texttt{F2} and \texttt{F3}, and \texttt{S1} is contained in the \texttt{\textbf{array}} \texttt{A0}, we say that \texttt{A0} is semantically dependent on \texttt{F0} (\texttt{F0}\ $\delta^s$\ \texttt{A0}).
          \end{itemize}

    \item \texttt{F2} is semantically dependent on \texttt{F0} (\texttt{F0}\ $\delta^s$\ \texttt{F2}).
\end{itemize}

\subsection{Semantic relations between fields.}
\begin{itemize}
    \item \texttt{F0} specifies the count of \texttt{A0};
          \texttt{A0.\textbf{count}} = \texttt{\textbf{int}(F0.\textbf{bytes}) - 1}.

    \item \texttt{F3} terminates \texttt{F2};
          \texttt{F2.\textbf{terminator} = F3.\textbf{bytes}}.

    \item \texttt{F4} terminates \texttt{F2};
          \texttt{F2.\textbf{terminator} = F4.\textbf{bytes}}.
\end{itemize}

%% file: appendix/bmp-use-case.tex
\section{Use case: BMP image}
\label{appendix:use-case-bmp}
The \texttt{bmp-parser}~\cite{bmp-parser-program} is a command-line utility tool that prints the file header, information header, and pixel array of a BMP image file. For this assessment, we used an uncompressed \texttt{5x2} BMP image file as an input to the \texttt{bmp-parser} program.

\begin{figure*}
    \centering
    \begin{minipage}{.47\textwidth}
        \centering
        \captionsetup{justification=centering, font=small}
        \includegraphics[scale=0.55]{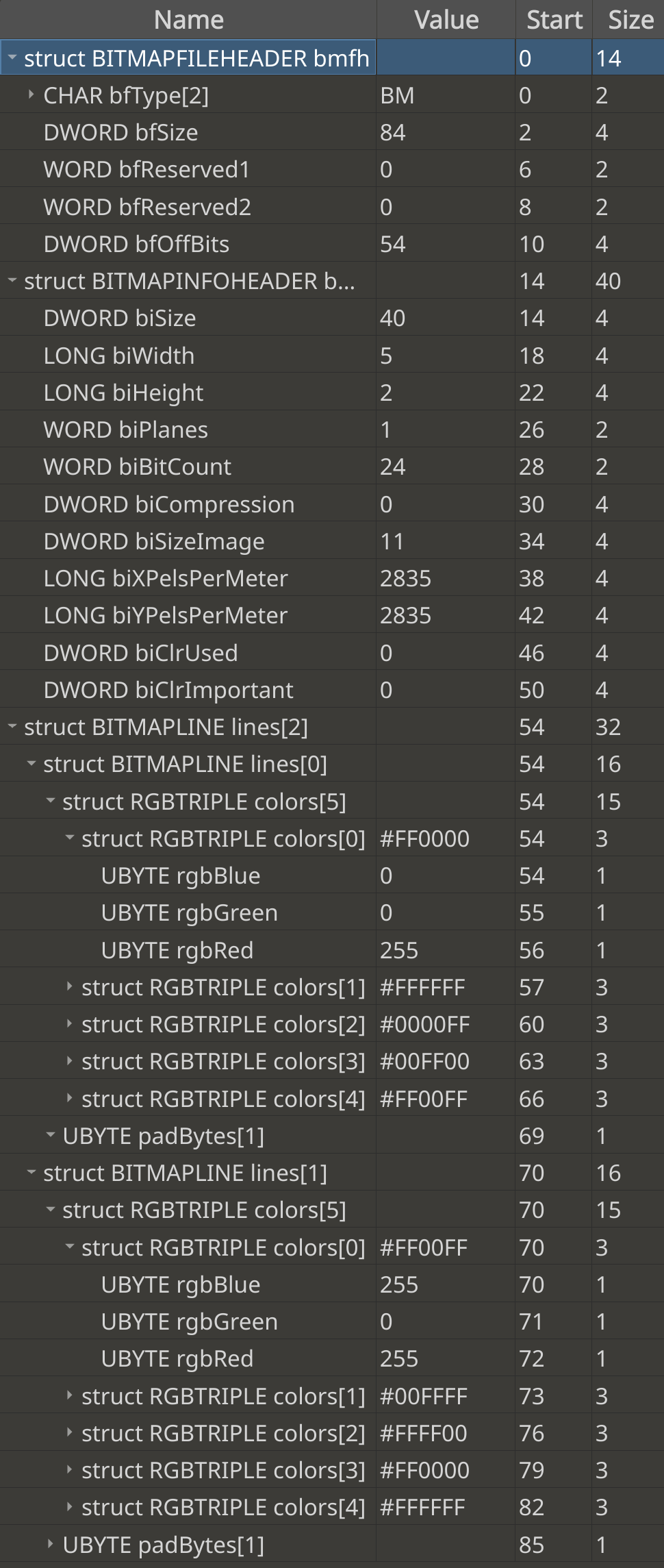}
        \captionof{figure}{Structure of a BMP image parsed by 010 Editor}
        \label{fig:bmp-010}
    \end{minipage}
    \begin{minipage}{.47\linewidth}
        \centering
        \captionsetup{justification=centering, font=small}
\begin{lstlisting}[
            style=structstyle,
            caption={Structure of BMP inferred by \byteriitwo},
            captionpos=b,
            label={lst:bmp_struct},
        ]
record S0 {
  atomic F0 (2) = <0x42,0x4d>,
  atomic F1 (4) = <0x54,0x00,0x00,0x00>;
  atomic F2 (4) = <0x00,0x00,0x00,0x00>;
  atomic F3 (4) = <0x36,0x00,0x00,0x00>;
  atomic F4 (4) = <0x28,0x00,0x00,0x00>;
  atomic F5 (4) = <0x05,0x00,0x00,0x00>;
  atomic F6 (4) = <0x02,0x00,0x00,0x00>;
  atomic F7 (2) = <0x01,0x00>;
  atomic F8 (2) = <0x18,0x00>;
  atomic F9 (4) = <0x00,0x00,0x00,0x00>;
  atomic F10 (12) = <0x0b,0x00,....0x00>;
  atomic F11 (4) = <0x00,0x00,0x00,0x00>;
  atomic F12 (4) = <0x00,0x00,0x00,0x00>;
  array A0 {
    record S1 {
      atomic F13 (1) = <ALL>,
      atomic F14 (1) = <ALL>;
      atomic F15 (1) = <ALL>;
    };
  },
    WHERE A0.count = int(F5.bytes) * int(F6.bytes),
    WHERE A0[0].F13.offset = int(F3.bytes);
}
\end{lstlisting}
    \end{minipage}
    \Description{}{}
\end{figure*}

The ground truth, that is, the BMP image file parsed by 010 Editor, is shown in Figure~\ref{fig:bmp-010} and \byteriitwo inferred structure is shown in Listing~\ref{lst:bmp_struct}.

As per the ground truth, there are three records in the given BMP image file: \texttt{BITMAPFILEHEADER} denoting BMP file header, \texttt{BITMAPFILEINFOHEADER} denoting BMP info header, and \texttt{\justify BITMAPFILELINE} and denoting bit map lines consisting of an array of pixels, \texttt{RGBTRIPLE}. Although \byteriitwo identifies the pixel array in the image file correctly, it overlooks separating the fields of the file header and info header into two different records. This is because \byteriitwo only groups a sequence of fields and turns them into a record if the same sequence of fields is repeated more than once in the input data. However, if we join the sequence of fields of the file header and info header identified by the ground truth and compare it with the corresponding sequence of fields in the structure identified by \byteriitwo, they are the same.

The mapping from the fields identified by 010 Editor to the corresponding fields in the structure inferred by \byteriitwo is as follows:
\begin{enumerate}
    \item Magic number: \texttt{bfType} $\rightarrow$ \texttt{F0},
    \item File size: \texttt{bfSize} $\rightarrow$ \texttt{F1},
    \item Reserved: \texttt{bfReserved1} $\rightarrow$ \texttt{F2},
    \item Reserved: \texttt{bfReserved2} $\rightarrow$ \texttt{F2},
    \item Offset of image data: \texttt{bfOffset} $\rightarrow$ \texttt{F3},
    \item Info header size: \texttt{biSize} $\rightarrow$ \texttt{F4},
    \item Image width: \texttt{biWidth} $\rightarrow$ \texttt{F5},
    \item Image height: \texttt{biHeight} $\rightarrow$ \texttt{F6},
    \item No of Planes: \texttt{biPlanes} $\rightarrow$ \texttt{F7},
    \item Bit count: \texttt{biBitCount} $\rightarrow$ \texttt{F8},
    \item Compression type: \texttt{\justify biCompression} $\rightarrow$ \texttt{F9},
    \item Image Size: \texttt{biSizeImage} $\rightarrow$ \texttt{F10},
    \item Horizontal resolution: \texttt{biXPelsPerMeter} $\rightarrow$ \texttt{F10},
    \item Vertical resolution: \texttt{biYPelsPerMeter} $\rightarrow$ \texttt{F10},
    \item No of Colors: \texttt{biClrUsed} $\rightarrow$ \texttt{F11},
    \item No of important colors: \texttt{biClrImportant} $\rightarrow$ \texttt{F12},
    \item Blue byte of a pixel: \texttt{rgbBlue} $\rightarrow$ \texttt{F13},
    \item Green byte of a pixel: \texttt{rgbGreen} $\rightarrow$ \texttt{F13}, and
    \item Red byte of a pixel: \texttt{rgbRed} $\rightarrow$ \texttt{F13}.
\end{enumerate}

{\justify Some of the consecutive fields identified by 010 Editor are grouped into a single field in the structure inferred by \byteriitwo. For instance, the two reserved fields in the file header, \texttt{bfReserved1} and \texttt{bfReserved2}, are grouped into a single field \texttt{F2} in \byteriitwo inferred structure. Similarly, three consecutive fields in the info header, \texttt{biSizeImage}, \texttt{biXPelsPerMeter}, and \texttt{biYPelsPerMeter}, are grouped into a single field \texttt{F10}. This is because the \texttt{bmp\_parse} program skips processing these fields, so the consecutive fields are grouped and a gap field is formed.}

The ground truth identifies two semantic relations:
\begin{enumerate}[(i)]
    \item \emph{Offset relationship.} \texttt{bfOffset} specifies the start offset of \texttt{RGBTRIPLE} array in the data section and
    \item \emph{Count relationship.} The total number of pixels in the image, that the number of elements of \texttt{RBGTRIPLE} array is the product of the \texttt{biWidth} and \texttt{biHeight}.
\end{enumerate}

\byteriitwo also identifies two semantic relations:
\begin{enumerate}[(i)]
    \item \emph{Offset relationship.} \texttt{A0.\textbf{offset}} = \texttt{F3.\textbf{bytes}}, where \texttt{A0} is the pixel array and \texttt{F3} is the offset field.
    \item \emph{Count relationship.} \texttt{A0.\textbf{count}} = \texttt{int(F5.\textbf{bytes}) *}\\ \texttt{int(F6.\textbf{bytes})}, where \texttt{F5} is the width field and \texttt{F6} is the height field.
\end{enumerate}

%% file: appendix/zip-use-case.tex
\section{Use case: ZIP file}
\label{appendix:use-case-zip}

\begin{figure*}
    \centering
    \begin{minipage}{.47\textwidth}
        \centering
        \captionsetup{justification=centering, font=small}
        \includegraphics[scale=0.60]{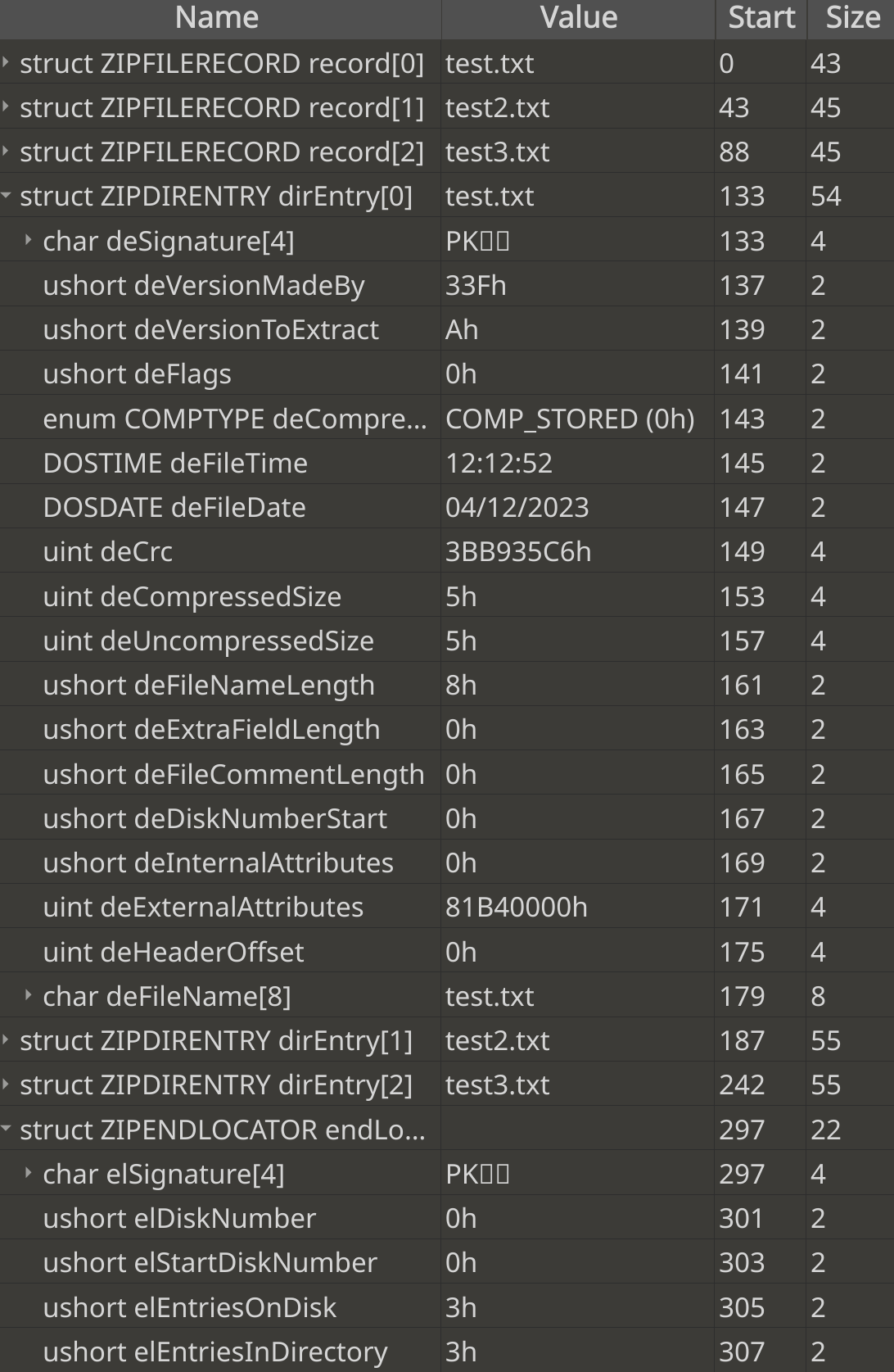}
        \captionof{figure}{Structure of a ZIP file parsed by 010 Editor}
        \label{fig:zip-010}
        \Description{}{}
    \end{minipage}
    \begin{minipage}{.47\linewidth}
        \centering
        \captionsetup{justification=centering, font=small}
\begin{lstlisting}[
            style=structstyle,
            caption={Structure of ZIP inferred by \byteriitwo},
            captionpos=b,
            label={lst:zip_struct},
        ]
record S0 {
  atomic F0 (133) = <0x50,0x4b,....0x0a>;
  array A0 {
    record S1 {
      atomic F1 (4) = <0x50,0x4b,0x01,0x02>;
      atomic F2 (1) = <0x3f>;
      atomic F3 (1) = <0x03>;
      atomic F4 (1) = <0x0a>;
      atomic F5 (1) = <0x00>;
      atomic F6 (2) = <0x00,0x00>;
      atomic F7 (2) = <0x00,0x00>;
      atomic F8 (4) = <CTRL,LOWER....0x56>;
      atomic F9 (4) = <ALL,ALNUM,ALL,PRINT>;
      atomic F10 (4) = <CTRL,0x00,0x00,0x00>;
      atomic F11 (4) = <CTRL,0x00,0x00,0x00>;
      atomic F12 (2) = <CTRL,0x00>;
      atomic F13 (2) = <0x00,0x00>;
      atomic F14 (2) = <0x00,0x00>;
      atomic F15 (2) = <0x00,0x00>;
      atomic F16 (2) = <0x00,0x00>;
      atomic F17 (4) = <0x00,0x00,0xb4,0x81>;
      atomic F18 (4) = <ALL,0x00,0x00,0x00>;
      atomic F19 (8+) = <0x74,0x65,....ER,+>,
         WHERE F19.size = int(F12.bytes);
    };
  };
  record S2 {
    atomic F20 (4) = <0x50,0x4b,0x05,0x06>;
    atomic F21 (4) = <0x00,0x00,0x00,0x00>;
    atomic F22 (2) = <0x03,0x00>;
    atomic F23 (2) = <0x03,0x00>;
    atomic F24 (8) = <0xa4,0x00,....0x00>;
    atomic F25 (2) = <0x00,0x00>;
  };
}
\end{lstlisting}
    \end{minipage}
\end{figure*}

A ZIP file parsed by 010 Editor is shown in Figure~\ref{fig:bmp-010} and \byteriitwo inferred structure by executing \texttt{zipinfo} with the same ZIP file is shown in Listing~\ref{lst:bmp_struct}. A ZIP file contains these sections:
\begin{itemize}
    \item An array of file records (\texttt{ZIPFILERECORD}), each containing content of files, compressed or uncompressed.
    \item An array of directory entries (\texttt{ZIPDIRENTRY}) records, each containing a file header of a file.
    \item A Trailer section (\texttt{ZIPENDLOCATOR}), which marks the end of a ZIP file.
\end{itemize}

\texttt{zipinfo} processes both the trailer section and the array of directory records and it skips reading file records of a ZIP file. This is why the array of file records was not recovered by \byteriitwo and it appears as a large data chunk (\texttt{F0}) in the structure. The parser in the 010 editor parses the file records as well.

\byteriitwo grouped directory entry records (\texttt{struct S1}) and created an array of directory records (\texttt{array A0}). It identified the trailer section as well (\texttt{struct S2}).

The mapping from the fields identified by the ground truth (\texttt{ZIPDIRENTRY}) to the corresponding fields in the directory entry structure identified by \byteriitwo (\texttt{struct S1}) is as follows:
\begin{enumerate}
    \item Directory signature: \texttt{deSignature} $\rightarrow$ \texttt{F1}
    \item Version made by: \texttt{deVersionMadeBy} $\rightarrow$ \texttt{F2, F3}
    \item Version to extract: \texttt{deVersionToExtract} $\rightarrow$ \texttt{F4, F5}
    \item Flags: \texttt{deFlags} $\rightarrow$ \texttt{F6}
    \item Compression type: \texttt{deCompression} $\rightarrow$ \texttt{F7}
    \item File last modification time: \texttt{deFileTime} $\rightarrow$ \texttt{F8}
    \item File last modification date: \texttt{deFileDate} $\rightarrow$ \texttt{F8}
    \item CRC of uncompressed data: \texttt{deCrc} $\rightarrow$ \texttt{F9}
    \item Compressed size: \texttt{deCompressedSize} $\rightarrow$ \texttt{F10}
    \item Uncompressed size: \texttt{deUnCompressedSize} $\rightarrow$ \texttt{F11}
    \item File name length: \texttt{deFileNameLength} $\rightarrow$ \texttt{F12}
    \item Extra field length: \texttt{deExtraFieldLength} $\rightarrow$ \texttt{F13}
    \item File comment length: \texttt{deFileCommentLength} $\rightarrow$ \texttt{F14}
    \item Disk \# where file starts: \texttt{deDiskNumberStart} $\rightarrow$ \texttt{F15}
    \item Internal attributes: \texttt{deInternalAttribute} $\rightarrow$ \texttt{F16}
    \item External attributes: \texttt{deExternalAttribute} $\rightarrow$ \texttt{F17}
    \item Header offset: \texttt{deHeaderOffset} $\rightarrow$ \texttt{F18}
    \item File Name: \texttt{deFileName} $\rightarrow$ \texttt{F19}
\end{enumerate}

The mapping from the fields identified by the ground truth (\texttt{ZIPENDLOCATOR}) to the corresponding fields in the trailer section structure identified by \byteriitwo (\texttt{struct S2}) is as follows:
\begin{enumerate}
    \item Trailer signature: \texttt{elSignature} $\rightarrow$ \texttt{F20}
    \item \# of this disk: \texttt{elDiskNumber} $\rightarrow$ \texttt{F21}
    \item Start disk \#: \texttt{elStartDiskNumber} $\rightarrow$ \texttt{F21}
    \item \# of directories in this disk: \texttt{elEntriesOnDisk} $\rightarrow$ \texttt{F22}
    \item Total number of directories: \texttt{elEntriesInDirectory} $\rightarrow$ \texttt{F23}
    \item Directory size: \texttt{elDirectorySize} $\rightarrow$  \texttt{F24}
    \item Directory offset: \texttt{elDirectoryOffset} $\rightarrow$  \texttt{F25}
    \item Comment Length: \texttt{elCommentLength} $\rightarrow$  \texttt{F25}
\end{enumerate}

\byteriitwo identifies that the length field \texttt{F12} specifies the length of the variable-length file name field \texttt{F19}.

%% file: appendix/png-use-case.tex
\section{Use case: PNG image}
\label{appendix:use-case-png}

\begin{figure*}
    \centering
    \begin{minipage}{.48\textwidth}
        \centering
        \captionsetup{justification=centering, font=small}
        \includegraphics[scale=0.25]{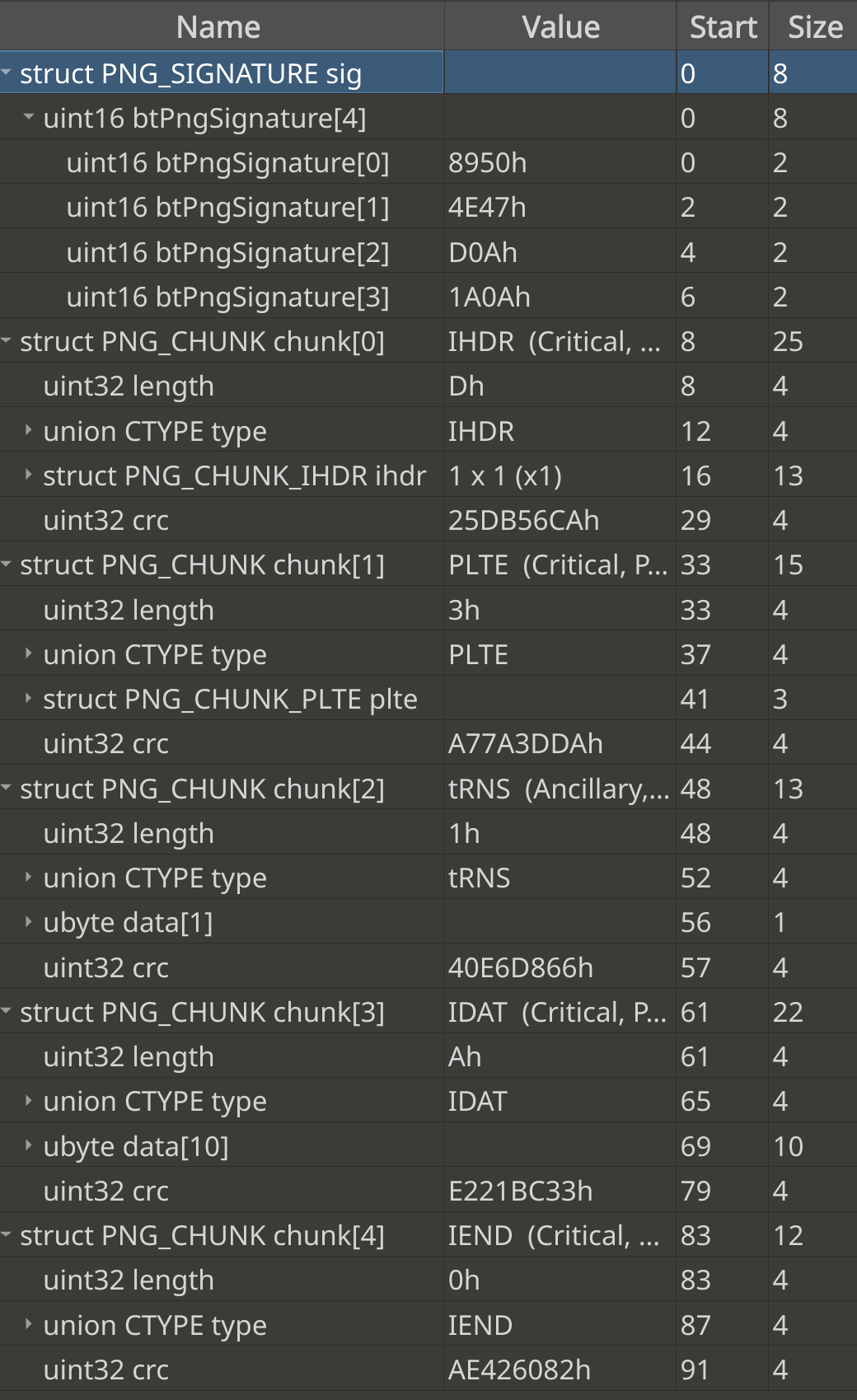}
        \captionof{figure}{Structure of a PNG file parsed by 010 Editor}
        \label{fig:png-010}
        \Description{}{}
    \end{minipage}
    \begin{minipage}{.48\linewidth}
        \centering
        \captionsetup{justification=centering, font=small}
\begin{lstlisting}[
            style=structstyle,
            caption={Structure of PNG inferred by \byteriitwo},
            label={lst:png_struct},
        ]
record S0 {
  atomic F0 (1) = <0x89>;
  array A0 {
    atomic F1 (1) = <UPPER>;
  };
  record S1 {
    atomic F2 (1) = <0x0d>;
    atomic F3 (1) = <0x0a>;
    atomic F4 (1) = <0x1a>;
    atomic F5 (1) = <0x0a>;
    array+ A1 {
      option U1 {
        record S2 {
          atomic F6 (4) = <0x00,0x00,0x00,WHITE>;
          atomic F7 (4) = <ALPHA,UPPE....PPER>;
          atomic F8 (8) = <0x00,0x00,....0x01>;
          atomic F9 (1) = <0x01>;
          atomic F10 (1) = <0x03>;
          atomic F11 (1) = <0x00>;
          atomic F12 (1) = <0x00>;
          atomic F13 (1) = <0x00>;
          atomic F14 (4) = <ALL,ALL,+>;
        };
        record S3 {
          atomic F6 (4) = <0x00,0x00,0x00,CTRL>;
          atomic F7 (4) = <ALPHA,UPPE....PPER>;
          atomic F15 (3) = <0x00,0x00,0x00>,
             WHERE F15.size = int(F6.bytes);
          atomic F14 (4) = <ALL,ALL,+>;
        };
        record S4 {
          atomic F6 (4) = <0x00,CTRL,+>;
          atomic F7 (4) = <ALPHA,UPPE....PPER>;
          atomic F16 (1) = <0x00>,
             WHERE F16.size = int(F6.bytes);
          atomic F14 (4) = <ALL,ALL,+>;
        };
        record S5 {
          atomic F6 (4) = <0x00,0x00,0x00,WHITE>;
          atomic F7 (4) = <ALPHA,UPPE....PPER>;
          atomic F18 (10) = <0x63,0x60,....0x01>;
             WHERE F18.size = int(F6.bytes);
          atomic F14 (4) = <ALL,ALL,+>;
        };
      };
    },
       WHERE A1.record_type = F7.bytes;
    record S6 {
      atomic F6 (4) = <0x00,0x00,0x00,CTRL>;
      atomic F7 (4) = <ALPHA,UPPE....PPER>;
      atomic F14 (4) = <ALL,ALL,+>;
    };
  };
}
\end{lstlisting}
    \end{minipage}
\end{figure*}

The PNG image file parsed by 010 Editor is shown in Figure~\ref{fig:bmp-010} and \byteriitwo inferred structure by executing \texttt{pngcheck} using the same image file is shown in Listing~\ref{lst:bmp_struct}. A PNG file contains these sections:
\begin{itemize}
    \item A file header (\texttt{PNG\_SIGNATURE}), which contains an 8-byte signature.
    \item An array of chunks (\texttt{PNG\_CHUNK}), each containing fields: \texttt{length}, \texttt{type}, \texttt{data}, and \texttt{data} fields.
\end{itemize}

010 Editor treats the signature as an array of 2-byte integers, whereas the \texttt{pngcheck} reads the first byte, the fifth byte to the eighth bytes separately, and it reads the second byte to the fourth byte in a loop.

The mapping from the fields identified by the ground truth (\texttt{PNG\_SIGNATURE}) to the corresponding fields of the PNG file signature identified by \byteriitwo is as follows:
\begin{enumerate}
    \item Highest bit set (\texttt{89}): \texttt{btPngSignature[0]} $\rightarrow$ \texttt{F0}
    \item ASCII letters PNG (\texttt{50 4E 47}): \texttt{btPngSignature[0]}, \texttt{btPngSignature[1]} $\rightarrow$ \texttt{F1}
    \item CLRF (\texttt{0D 0A}): \texttt{btPngSignature[1], btPngSignature[2]} $\rightarrow$ \texttt{F2, F3}
    \item EOF (\texttt{1A}): \texttt{btPngSignature[2]} $\rightarrow$ \texttt{F4}
    \item LF (\texttt{0A}): \texttt{btPngSignature[2]} $\rightarrow$ \texttt{F5}
\end{enumerate}

The mapping from the fields identified by the ground truth (\texttt{PNG\_CHUNK}) to the corresponding fields of the array of chunks structures (\texttt{S2-S6}) identified by \byteriitwo is as follows:
\begin{enumerate}
    \item Length of Data: \texttt{length} $\rightarrow$ \texttt{F6}
    \item Type of the record: \texttt{CTYPE type} $\rightarrow$ \texttt{F7}
    \item Data: \texttt{data}
    \item CRC value of Data: \texttt{crc} $\rightarrow$ \texttt{F14}
\end{enumerate}

Struct \texttt{S6} in Listing~\ref{lst:png_struct} is the IEND chunk.

\byteriitwo identifies that field \texttt{F7} specifies the type of record in each chunk. It also identifies the length relationship between the length field (\texttt{F6}) and the data field (\texttt{F16}).